\documentclass{aa}
\usepackage[varg]{txfonts}
\usepackage{graphicx,subcaption}
\usepackage{natbib}
\usepackage{rotating}
\usepackage{pdflscape}
\usepackage[colorinlistoftodos]{todonotes}
    
\begin{document}

\title{The ALMA Frontier Fields Survey}

\subtitle{II. Multiwavelength Photometric analysis of 1.1mm continuum sources in Abell 2744, MACSJ0416.1-2403 and MACSJ1149.5+2223}

\author{N. Laporte\inst{1,2}
     \and F. E. Bauer\inst{2,3,4}
     \and P. Troncoso-Iribarren\inst{2,23}
     \and X. Huang\inst{5,6}
     \and J. Gonz{\'{a}}lez-L{\'{o}}pez\inst{2}
     \and {S. Kim}\inst{2}
     \and {T. Anguita}\inst{7,3}
     \and {M. Aravena}\inst{8}
     \and {L. F. Barrientos}\inst{2}
     \and {R. Bouwens}\inst{20}
     \and {L. Bradley}\inst{22}
     \and {G. Brammer}\inst{22}
     \and {M. Carrasco}\inst{14}
     \and {R. Carvajal}\inst{2}
     \and {D. Coe}\inst{22}
     \and {R. Demarco}\inst{12}
     \and {R. S. Ellis}\inst{1,18}
     \and {H. Ford}\inst{15}
     \and {H. Francke}\inst{2}
     \and {E. Ibar}\inst{14}
     \and {L. Infante}\inst{2}
     \and {R. Kneissl}\inst{9.10}
     \and {A. M. Koekemoer}\inst{21}
     \and {H. Messias}\inst{20}
     \and {A. Mu{\~{n}}oz Arancibia}\inst{14}
     \and {N. Nagar}\inst{12}
     \and {N. Padilla}\inst{2}
     \and {R. Pell\'o}\inst{16}
     \and {M. Postman}\inst{21}
     \and {D. Qu\'enard}\inst{17}
     \and {C. Romero-Ca\~nizales}\inst{3,8}
     \and {E. Treister}\inst{2,12}
     \and {E. Villard}\inst{9,10}
     \and {W. Zheng}\inst{15}
     \and {A. Zitrin}\inst{11}}

\institute{ Department of Physics and Astronomy, University College London, Gower Street, London WC1E 6BT, UK
  \and Instituto de Astrof{\'{\i}}sica and Centro de Astroingenier{\'{\i}}a, Pontificia Universidad Cat{\'{o}}lica de Chile, Casilla 306, Santiago 22, Chile
         \and
  {Millennium Institute of Astrophysics, Chile} 
         \and
{Space Science Institute, 4750 Walnut Street, Suite 205, Boulder, Colorado 80301} 
         \and CAS Key Laboratory for Research in Galaxies and Cosmology, University of Science and Technology of China
         \and Department of Physics and Astronomy, Johns Hopkins University, 3400 N. Charles Street, Baltimore, MD 21218
         \and
{Departamento de Ciencias F\'{i}sicas, Universidad Andres Bello, 252 Avenida Rep\'{u}blica, Santiago, Chile}
		 \and
{N\'ucleo de Astronom\'{\i}a de la Facultad de Ingenier\'{\i}a y Ciencias, 
Universidad Diego Portales, Av. Ej\'ercito 441, Santiago, Chile}
		 \and
{Joint ALMA Observatory, Alonso de C\'{o}rdova 3107, Vitacura, Santiago, Chile}
         \and
{European Southern Observatory, Alonso de C\'{o}rdova 3107, Vitacura, Casilla 19001, Santiago, Chile}
         \and
{Physics Department, Ben-Gurion University of the Negev, P.O. Box 653, Be’er-Sheva 8410501, Israel}
		 \and
{Department of Astronomy, Universidad de Concepcion, Casilla 160-C, Concepci\'{o}n, Chile}
		  \and
{Zentrum f\"{u}r Astronomie, Institut f\"{u}r Theoretische Astrophysik, Philosophenweg 12,
69120 Heidelberg, Germany}
		 \and
{Instituto de F\'isica y Astronom\'ia, Universidad de Valpara\'iso, Avda. Gran Breta\~na 1111, Valpara\'iso, Chile}
		 \and
{Department of Physics and Astronomy, Johns Hopkins University, Baltimore, MD 21218, USA}
         \and
{Institut de Recherche en Astrophysique et Planétologie (IRAP), Université de Toulouse, CNRS, UPS, 14 Av. Edouard Belin, F-31400 Toulouse, France}
        \and
{School of Physics and Astronomy Queen Mary, University of London, 327 Mile End Road, London E1 4NS, UK}
		\and
{European Southern Observatory (ESO), Karl-Schwarzschild-Strasse 2, 85748 Garching, Germany}
       \and
{Leiden Observatory, Leiden University, NL-2300 RA Leiden, Netherlands}
		\and
{Instituto de Astrof\'isica e Ci\^encias do Espa\c co, Universidade de Lisboa, OAL, Tapada da Ajuda, PT 1349-018 Lisboa, Portugal}
       \and
{Space Telescope Science Institute, Baltimore, MD, USA}
         \and
{Departamento de Ciencias F\'{i}sicas, Universidad Andres Bello, 252 Avenida Rep\'{u}blica, Santiago, Chile}
		\and 
{Universidad Autonoma, Av. Pedro de Valdivia 425, Santiago, Chile}
}

\date{Received  / Accepted }

 \abstract
   { The \textit{Hubble} and {\it Spitzer} Space Telescope surveys of the Frontier Fields provide extremely deep images around six massive, strong-lensing clusters of galaxies. The ALMA Frontier Fields survey aims to cover the same fields at 1.1mm, with maps reaching (unlensed) sensitivities of $<$70\,$\mu$Jy, in order to explore the properties of background dusty star-forming galaxies.}
   {We report on the multi-wavelength photometric analysis of all 12 significantly detected ($>$5$\sigma$) sources in the first three \textit{Frontier Fields} clusters observed by ALMA, based on data from \textit{Hubble} and \textit{Spitzer}, the Very Large Telescope and the \textit{Herschel} Space Observatory.}
   {We measure the total photometry in all available bands and determine the photometric redshifts and the physical properties of the counterparts via SED-fitting. In particular, we carefully estimate the far-infrared (FIR) photometry using 1.1\,mm priors to limit the misidentification of blended FIR counterparts, which strongly affect some flux estimates in previous FIR catalogs. Due to the extremely red nature of these objects, we used a large range of parameters (e.g. 0.0 $< A_v <$20.0) and templates (including AGNs and ULIRGs models). }
   { We identify robust near-infrared (NIR) counterparts for all 11 sources with K$_s$ detection, the majority of which are quite red, with eight having $F814W-K_s\gtrsim 4$ and five having $F160W-[4.5]\gtrsim3$.From the FIR point of view, all our objects have $z_{phot}$$\sim$1--3, whereas based on the optical SED one object prefers a high-$z$ solution ($z\geq\ $7). Five objects among our sample have spectroscopic redshifts from the GLASS survey for which we can reproduce their SEDs with existing templates. This verification confirms the validity of our  photometric redshift methodology. The mean redshift of our sample is $z_{phot}$=1.99$\pm$0.27. All 1.1\,mm selected objects are massive (10.0$<\log[M_{\star}(M_{\odot})]<$ 11.5), with high star formation rates ($<\log[SFR(M_{\odot}/yr)]> \approx$1.6) and high dust contents (8.1 $<\log[M_{dust} (M_{\odot})]<$8.8), consistent with previous ALMA surveys. }
   {}
\maketitle

\section{Introduction}

    
In 2013, the \textit{Hubble} Space Telescope ({\textit{HST}) initiated the \textit{Frontier Fields} survey \citep{2017ApJ...837...97L}, with observations of six massive clusters to unprecedented depths, with the goal of improving our understanding of faint galaxies in the high-$z$ Universe. Thanks to this new legacy program, the number of $z$$>$6.5 candidates has significantly increased (\citealt{2015ApJ...800...18A}, \citealt{2015ApJ...799...12I}, \citealt{2014A&A...562L...8L}, \citealt{2015A&A...575A..92L}), allowing tighter constraints on the physical properties of the first galaxies (\citealt{2015ApJ...804..103K}, \citealt{2017ApJ...837L..21L}) up to very high-redshifts (\citealt{2015arXiv151007084I}, \citealt{2014ApJ...793L..12Z}). The exquisite data allow constraints on many types of $z$$\sim$1--3 galaxies as well. The ALMA \textit{Frontier Fields} survey (ID 2013.1.00999.S, PI: F. Bauer) was designed to produce deep $\approx$2\farcm1$\times$2\farcm2 maps at 1.1mm covering the \textit{HST}/WFC3 fields-of-view for all six clusters. In the first phase, we have produced these maps for Abell 2744 (hereafter A2744), MACSJ0416.1-2403 (hereafter M0416) and MACSJ1149.5+2223 (hereafter M1149), with unlensed sensitivities of 55, 59 and 71 $\mu$Jy/beam respectively. Recently \citet{Gonzalez2016} reported a list of twelve 1.1\,mm continuum detections in these three clusters. We report here on the photometric analysis of all 1.1\,mm continuum detections, combining data from the \textit{HST}, VLT, \textit{Spitzer} and \textit{Herschel} observatories. 

Our aim is to place these 12 ALMA-FF detections in context compared to the brighter submillimeter (submm) and FIR-detected sources that have been extensively studied to date \citep[e.g.,][]{1997ApJ...490L...5S, 1998Natur.394..248B, 2012A&A...539A.155M, 2013Natur.495..344V, 2014ApJ...788..125S, 2015ApJ...799..194C} and ultimately to understand what role dusty star-forming galaxies (DSFGs) play in the evolution of massive objects over cosmic time \citep[e.g.,][]{2014PhR...541...45C}. 

In section \ref{sec.data} we present the properties of all datasets used to constrain the spectral energy distributions (SEDs) of the ALMA sources. We explain our search for optical/NIR counterparts in section \ref{sec.counterparts}. The extraction of their photometry is described in section \ref{sec.photometry}. The physical properties of these objects, including their photometric redshifts, reddening and star formation rates, are reported in section \ref{section.sed}. 
Throughout this paper, we use a concordance cosmology ($\Omega_M = 0.3$, $\Omega_{\Lambda} = 0.7$ and $H_0 = 70$ km/s/Mpc), all magnitudes are quoted in the AB system \citep{1983ApJ...266..713O} and all significances refer to reduced values of $\chi_{\nu}^2$ .

\section{Multi-wavelength Data}
\label{sec.data}

In this section, we describe all the data -- from the \textit{Hubble}, VLT, \textit{Spitzer}, \textit{Herschel} observatories -- that were used in the analysis of the ALMA detected sources. 

We used ACS $F435W$, $F606W$, $F814W$ and WFC3 $F105W$, $F125W$, $F140W$, $F160W$-filter images obtained within the framework of the \textit{Frontier Fields} (\textit{FFs}) legacy survey and reduced by the \textit{Space Telescope Science Institute}. All \textit{HST} survey data acquired in these fields (IDs 14041 PI: P. Kelly; 13495, 13496, 13504 PI: J. Lotz; 13386 PI: S. Rodney; 12459 PI: M. Postman; 11689 PI: R. Dupke) were retrieved from the STScI Mikulski Archive for Space Telescopes (MAST), where we use the final full-depth v1.0 mosaics that were produced by the Frontier Fields Team at STScI, using the latest calibration files for each cluster\footnote{https://archive.stsci.edu/prepds/frontier/}. Limiting magnitudes were estimated from 0\farcs4 radius apertures distributed all over the field. 

We took benefit of the deep $K_{s}$ images obtained with HAWK-I/VLT \citep{2008A&A...491..941K} around A2744 and M0416 as part of ESO program 092.A-0472 (PI: G. Brammer, \citealt{2016arXiv160607450B}). We estimated limiting magnitudes from the rms measured in 0\farcs4 apertures distributed over the field. We applied aperture corrections of 1.48 and 1.41, respectively for the A2744 and M0416 images \citep{2016ApJS..226....6B}.

We used images acquired with the IRAC and MIPS instruments onboard the \textit{Spitzer Space Telescope} covering wavelengths from $\approx$ 3.6--24\,$\mu$m for A2744 and 3.6--4.5\,$\mu$m for M0416 and M1149 (where no 5.8, 8.0 and 24 $\mu$m data are available). The reduction of the A2744 and M0416 images at 3.6 and 4.5 $\mu$m are described in detail in \citet{2014A&A...562L...8L} and \citet{2015A&A...575A..92L}, while further details on the reduction of IRAC data at 3.6 and 4.5 $\mu$m for M1149 will be provided in Zheng et al. (in prep.). We additionally used the public SEIP Super Mosaic images from the {\it Spitzer} Heritage Archive at 5.8, 8.0 and 24$\mu$m for A2744. We measured 5$\sigma$ limiting depths using 1\farcs2 radius apertures distributed over the entirety of the blank sky in each image. 

All three clusters were also observed by the \textit{Herschel} Space Observatory \citep{2010A&A...518L...1P} within the framework of the \textit{Herschel} Lensing Survey\footnote{http://herschel.as.arizona.edu/hls/hls.html} \citep[HLS; IDs KPOT\_eegami\_1, OT2\_eegami\_5;][]{2010A&A...518L..12E}. We used the publicly available images reduced by HLS for A2744 and M1149, and the level 2.5 PACS and level 3 SPIRE maps processed by the \textit{Herschel} Science Centre for M0416, to add photometric constraints on the SEDs of the ALMA detected sources between $\approx$85--600\,$\mu$m. We measured the 5$\sigma$ depths of these images using apertures set to the beam size in each band (see Table~1). We applied aperture corrections to the SPIRE photometry assuming spectral indices ranging from -4 to 4 and using values tabulated in the SPIRE Handbook.

The properties of our dataset are summarized in Table~\ref{tab.data_properties}.


\begin{table*}
\caption{Properties of the multi-wavelength dataset.}             
\label{tab.data_properties}      
\centering                          
\begin{tabular}{c | c c | c c c | c | c }        
\hline\hline                 
Filter & $\lambda_c$ & $\delta \lambda$ & A2744 & M0416 & M1149 & Instrument & Aperture  \\    
         &  [$\mu$m]  & [$\mu$m]  & [AB]  &  [AB]  & [AB] &  & [''] \\          
 \hline                        
F435W        &  0.431  &  0.073    &  28.8  &  28.7  &  28.4  &  ACS & 0.4 \\
F606W        &  0.589  &  0.157  &  28.8  &  28.9  &  28.7  &  ACS  & 0.4 \\
F814W        &  0.811  &  0.166  &  29.1  &  29.2  &  28.9  &  ACS  & 0.4 \\
F105W        &  1.050  &  0.300  &  29.6  &  29.6  &  29.5  & WFC3  & 0.4 \\
F125W        &  1.250  &  0.300  &  29.4  &  29.3  &  29.4  & WFC3  & 0.4 \\
F140W        &  1.400  &  0.400  &  29.4  &  29.3  &  29.2  & WFC3  & 0.4 \\
F160W        &  1.545  &  0.290  &  29.3  &  29.3  &  29.4  & WFC3 &  0.4 \\ \hline
K$_s$         &  2.146  &  0.324  &  26.2  &  26.3  & -         & HAWK-I  & 0.4 \\ \hline
3.6$\mu$m &  3.550   &  0.743  &  25.1  &  25.6  & 25.0  & IRAC  & 1.2\\
4.5$\mu$m &  4.493  &  1.010 &  25.3  &  25.7  &  25.0  & IRAC  & 1.2 \\
5.8$\mu$m &  5.738  &  1.256  & 22.7  &  -        & -        & IRAC  & 1.2 \\
8.0$\mu$m &  7.927  &  2.831  & 22.6  & -         &  -       & IRAC  & 1.2 \\
24$\mu$m  & 23.843 & 53.245 & 18.8  & -         & -        & MIPS  & 7.5 \\ \hline
Blue            & 71.933 & 22.097 & -        & -         & 13.4  & PACS  & 5.2 \\
Green         & 102.62 & 35.686 & 15.1  & 14.8  & 13.8  & PACS   & 7.7 \\
Red            &  167.13 & 74.954 & 13.4 & 14.1  & 13.6   & PACS  & 12 \\
PSW           &  251.50 & 67.615 & 13.9 & 13.3  & 13.9 & SPIRE  & 22 \\
PMW          &  352.83 & 95.756 &  13.4 & 13.5  & 13.5 & SPIRE & 30 \\
PLM           &  511.61  & 185.672 & 13.4 & 13.4  & 13.4 & SPIRE  & 42 \\
\hline
\end{tabular}

\vspace{0.6cm}
Columns: (1) Filter ID, (2) Central Wavelength, (3) FWHM, (4,5,6) 5$\sigma$ limiting magnitude for the 3 clusters, (7) Instrument, (8) Aperture radius in which the depth is measured \\
\end{table*}


\section{Search for NIR-counterparts}
\label{sec.counterparts}

The identification of optical-NIR counterparts to bright submm/mm sources has been extensively discussed (e.g.   \citealt{2002MNRAS.331..495S}, \citealt{2004ApJS..154..137F}, \citealt{2010MNRAS.409L..13C}, \citealt{2011MNRAS.416..857S}). Thanks to the ALMA beam size, the number of possible counterparts is strongly reduced compared to typical single-dish bolometer surveys. For the ALMA-FF survey, we principally searched for NIR counterparts using the deep \textit{HST} $F160W$ images, since the 1.1\,mm sources are likely to lie at $z$$>$1 and be relatively red. We examined all the sources residing within a circle centered on the ALMA position with a radius of 2$\times$ the ALMA average FWHM of the synthesized beam (i.e., 1\farcs1--2\farcs3) as measured in \citet{Gonzalez2016}; these values were chosen to account for potential offsets between stellar and dust emission \citep[e.g.,][]{2013ApJ...776...22H, 2014ApJ...785..111W}. 

We refer the reader to Fig.~11 of \citet{Gonzalez2016}, where the positional offsets are already presented. Remarkably, for all but one of the $>$5$\sigma$ sources, there is a clear near-IR counterpart within $\lesssim$0\farcs2 of the ALMA position. Moreover all these counterparts have characteristically red colors, as might be expected for DSFGs selected from ALMA maps \citep[e.g.,][]{2002MNRAS.331..495S, 2004ApJ...606..664D, 2012ApJ...744..155W, 2016ApJ...820...82C}. It is more difficult to determine an NIR counterpart for A2744-ID02, which is not centered on a strong NIR source. Two faint NIR objects formally reside inside the 1\farcs1 search circle, although neither is well aligned with the ALMA position, which appears to lie in between the counterparts. Both A2744-ID02 candidates have red colors, similar to the rest of the ALMA-FF sources. Intriguingly, there is faint diffuse $F160W$ emission extending between the two NIR counterparts, and the resolved ALMA emission appears to be elongated roughly coincident with a suppression in the $F160W$ emission. This suggests that the ALMA source may arise from a dusty region that divides these two $F160W$ sources, which may represent less-obscured clumps from a single extended object with variable and strong extinction (see Fig.~\ref{fig.stamps.gold}). Interestingly, in the $K_{s}$ and IRAC 3.6--8.0\,$\mu$m images, the flux at the ALMA position increases relative to the two $F160W$ sources as wavelength increases, such that by 8\,$\mu$m the peak emission is in fact centered almost exactly on the ALMA position. For the purposes of SED-fitting, we adopt the nearest NIR counterpart for A2744-ID02, but caution that the true SED at the ALMA position may suffer significantly stronger extinction. 

The optical-NIR counterparts for all the ALMA $>$5$\sigma$ detections are displayed on Fig.~\ref{fig.stamps.gold}. 

One interesting point to consider with regard to the positional offsets, first noted by \citet{2014ApJ...785..111W}, is that the central positions of the submm and optical/NIR emission may not coincide due to strong dust extinction, an  effect which should increase with increasing redshift since bluer emission is more easily extincted. While A2744-ID02 is a rather obvious case, in fact we find that such offsets are present for a large majority of  the ALMA-FFs sample, where in nine out of 12 cases (i.e., A2744-ID01, A2744-ID02, A2744-ID03, A2744-ID04, A2744-ID06, A2744-ID07, M0416-ID02, M0416-ID03, and M0416-ID04) the ALMA centroid position falls on a relatively darker region of the counterpart galaxy (see Fig.~\ref{fig.stamps.gold}, as well as Fig.~11 and Table~6 of \citealt{Gonzalez2016}). This physical effect is likely the dominating term in the measured offsets.

\begin{figure*}
\centering
\includegraphics[width=18.5cm]{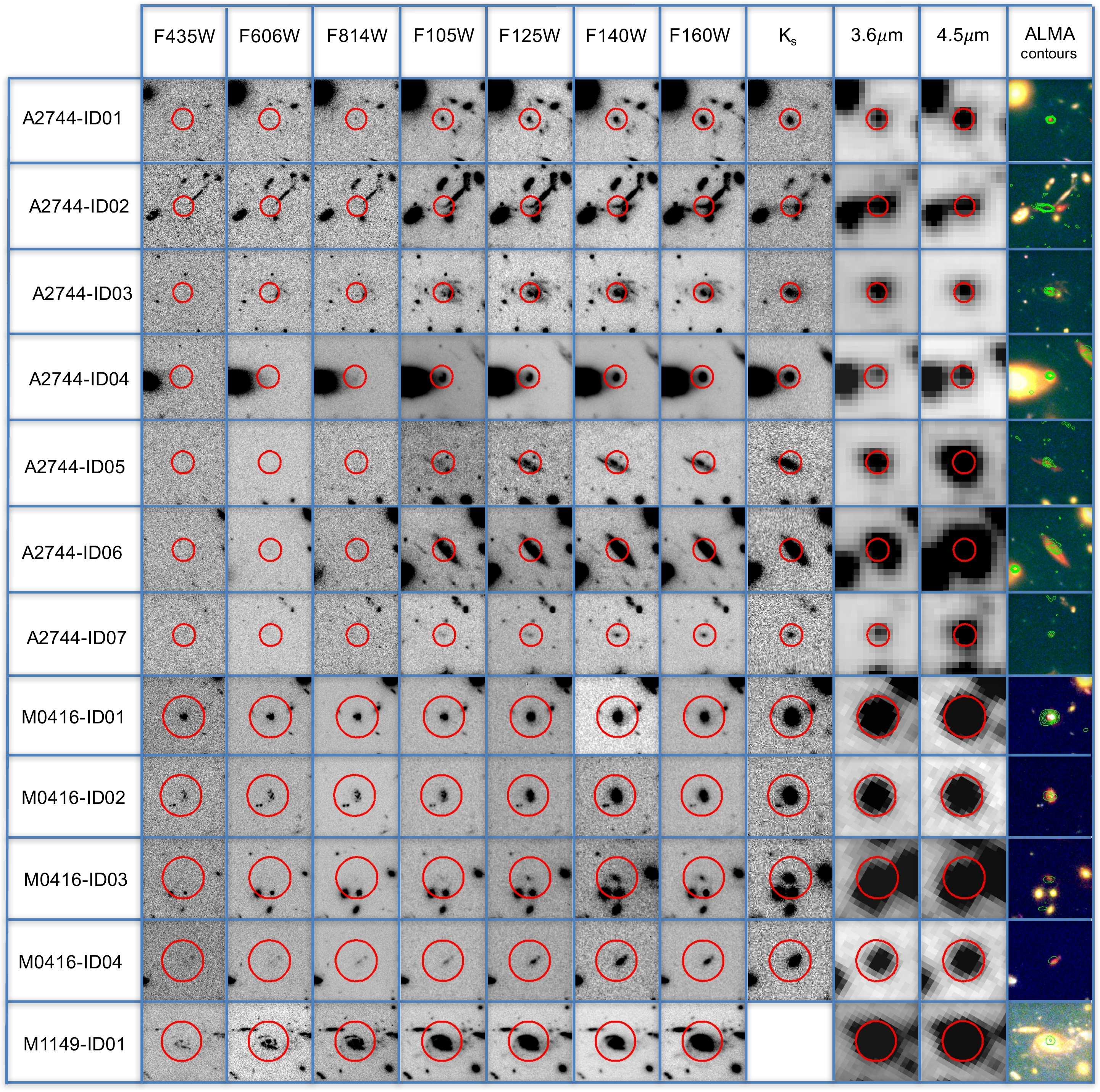}
\caption{\label{fig.stamps.gold}Thumbnail images of the optical/NIR counterpart for each ALMA $>$5$\sigma$ source. Image sizes are 9\arcsec$\times$9\arcsec. From left to right, the columns denote {\it HST} filters $F435W$, $F606W$, $F814W$, $F105W$, $F125W$, $F140W$, and $F160W$, {\it Spitzer} IRAC filters 3.6\,$\mu$m and 4.5\,$\mu$m, and lastly a color image composed of {\it HST} filters $F814W$ (blue), $F105W$ (green) and $F160W$ (red), with green contours denoting 1.1mm emission detection at 4, 6, 8, 10, and 12 $\sigma$. Red circles are centered on the ALMA centroid position and denote the search radius employed (1\farcs1 for A2744, and 2\farcs3 for M0416 and M1149).
As described in the text, A2744-ID02 has multiple counterparts, we used the counterpart highlighted by the red circle to estimate its SED}
\end{figure*}

\section{Photometry}
\label{sec.photometry}

The photometry of the identified counterparts from the optical (\textit{HST}) to the FIR (\textit{Herschel}) was estimated as follows.

\subsection{Hubble Space Telescope}
\label{HST.sec}

In 2016, the \textit{AstroDeep} project released a first version of photometric catalogs based on Frontier Fields images of A2744 and M0416 obtained with ACS/HST, WFC3/HST, HAWKI/VLT and IRAC/Spitzer (\citealt{2016arXiv160302460M}, \citealt{2016arXiv160302461C}). In their analysis, they remove foreground emission from intra-cluster light and bright galaxies. Most of our targets, excepted A2744-ID02, are in these catalogs and, in the following, we will use the HST photometry they extracted for each source. However, for a few objects, our visual inspection demonstrates that some detections or non-detections in ACS images are not correct in the \textit{AstroDeep} catalog. For example: A2744-ID04 appears undetected to $>$27.8AB in F606W while \textit{AstroDeep} has a detection of 24.9AB; A2744-ID06 is detected in F606W at 26.75AB but undetected in \textit{AstroDeep}; A2744-ID07 is not detected in F606W at $>$29.8AB yet has detection in \textit{AstroDeep}; and M0416-ID03 is clearly not detected at F606W to $>$29.2AB while an \textit{AstroDeep} detection of $m_{F606W} \sim$26.8 is claimed (see Fig.~\ref{fig.stamps.gold}). For all the objects, we updated the detection/non-detection according to our visual inspection. 


For the two sources that are not included in AstroDeep catalogs, A2744-ID02 and M1149-ID01, we used SExtractor v2.19.5 \citep{1996A&AS..117..393B} in double image mode on PSF-matched \textit{HST} data using a sum of WFC3 data as the detection image. We set the extraction parameters, taking into account the FWHMs of the ACS and WFC3 images, as follows:

\begin{itemize}
\item DETECT\_MINAREA : 5 pixels above the threshold
\item DETECT\_THRESHOLD : 1.5$\sigma$
\item DEBLEND\_NTHRESH : 32 deblending sub-thresholds
\item DEBLEND\_MINCONT : 0.005 (contrast for deblending)
\end{itemize}
The number of detections per field is $\sim$7000, 9000 and 16000, respectively, for A2744, M0416 and M1149 over the $\approx$2\arcmin $\times$2\arcmin\ field of view. 
We estimate aperture corrections by comparing the flux measured in a Kron aperture defined by a Kron factor of 1.2 and a minimum radius of 1.7 with the SExtractor MAG\_AUTO \citep{2007ApJ...670..928B} and averaged for $\approx$30 point-like objects distributed over the field in each band. Error bars are estimated from the RMS measured in several apertures distributed over the field and take into account uncertainties on the zero-point. 


\subsection{Ground-based Telescopes}

The photometry of our objects in the deep $K_{s}$ images obtained with HAWK-I/VLT was measured using 0\farcs4 radius apertures with the IRAF NOAO {\tt daophot} package. In order to estimate the total flux belonging to our objects, we applied aperture corrections as described in \citet{2016arXiv160607450B}. We computed error bars based on the blank-sky noise measured in the vicinity of each object using the same sized aperture. 

\subsection{Spitzer Space Telescope}

To complement the \textit{HST} dataset, deep \textit{Spitzer}/IRAC channel 1 and 2 images, which correspond to 3.1--3.9 and 3.9--5.0\,$\mu$m, respectively, were acquired for A2744 (Zheng et al. 2014), M0416 (Infante et al. 2015) and M1149 (Zheng et al. 2016, in preparation). We also analyzed the IRAC channel 3 and channel 4 data of A2744 , which correspond to bandwidths of 4.8--6.5 and 6.2--9.3$\mu$m, respectively. The IRAC images of our candidates suffer from crowding in some cases due to the instrument’s large point spread function (PSF, FWHM $\sim$1\farcs9--2\farcs0), such that simple aperture photometry occasionally results in inaccurate fluxes due to contamination from nearby sources. To address this issue, we adopt a deblending technique with the help of the GALFIT software \citep{2010AJ....139.2097P}. In this method, we perform a fit to the objects of interest and all their nearby neighbors simultaneously in a $\sim$10\arcsec$\times$10\arcsec\ fitting window around the target. All the sources falling in this window are fitted with PSF models or Sersic models when necessary. The PSF is determined from the same IRAC image using several nearby bright, isolated point sources. The initial positions and profiles of each model source are derived from the higher resolution \textit{HST} $F160W$-band mosaic images. During the fitting process, all input parameters are allowed to vary, while the relative positions of the objects are tied together.


We also used GALFIT to extract fluxes for our objects in the MIPS 24$\mu$m image of A2744. We modeled and removed all the nearby sources assuming a Sersic profile, and then measured the remaining flux for each object in the residual map in a 7\farcs5 aperture. We then applied aperture corrections to obtain total fluxes according to the MIPS User Manual.
    
Finally, we compared the colors and photometry measured here for all 1.1mm sources detected on ALMA maps of A2744 and M0416 against those from the AstroDeep catalogs. We find consistency for nearly all measurement within the errors (excepted for the ACS detection/non-detection highlighted in section 4.1) , confirming the method we used to extract \textit{HST}, HAWK-I and \textit{Spitzer} photometry.
    
\subsection{Herschel Space Observatory}

The data from PACS (100/160 $\mu$m) and SPIRE (250/350/500 $\mu$m) for all three clusters were taken in the framework of the HLS \citep{2010A&A...518L..12E}. However, HLS only provides publicly reduced data for A2744 and M1149, so level 2.5 PACS and level 3 SPIRE images for M0416 were obtained from the {\it Herschel} archive. The astrometry of the {\textit{Herschel} images was fixed to match the point sources in the IRAC and MIPS images for A2744, and the IRAC images for M0416 and M1149.

Because of the large beam sizes of the \textit{Herschel} instruments, several blended sources can account for the emission observed within one beam, making it difficult to measure the true flux density of a given galaxy. 

The photometry was thus obtained as follows: the positions of the ALMA detected sources were used as priors for the total emission in the PACS and SPIRE images. In a few cases, additional priors were required to account for all of the emission in the PACS bands, based on bright IRAC or MIPS sources that were added by hand. This assumption is reasonable since at longer wavelengths, where the blending is higher, the 1.1\,mm emission should be a good indicator of which galaxies are responsible for the far-IR emission. At the same time, at the shorter wavelengths, the 1.1\,mm emission might not provide as reliable a guide. This effect is contrasted by the fact that at shorter wavelengths the blending problems are less important since the beam sizes are smaller. 





The flux density was measured by fitting the observed emission in each image with a set of point sources (modeled with the beam response) located at the positions of the priors. The flux corresponding to each point source was left to vary following an MCMC sampler. The observed emission was fitted by all the sources at the same time. This step was critical to sample the degeneracy and associated uncertainties produced when emission from nearby FIR-bright sources fell within the beam of the ALMA sources. The best solutions were obtained by using maximum likelihood estimations with the provided uncertainties as Gaussian errors. The flux densities for the sources were obtained from the posterior probability distribution of the fluxes for each of the point sources used in the fitting procedure. The quoted flux densities correspond to the median value of the distribution and the errors encompass the 1$\sigma$ range. Sources with flux density values lower than 3 times the measured sigma range are considered non-detections and 3$\sigma$ upper limits are provided instead. In some cases, the 1$\sigma$ values obtained from the posterior probability distribution were smaller than the rms uncertainties found from blank-sky regions of the images. In such cases, the errors were obtained by combining the aforementioned uncertainties in quadrature.

A thorough investigation of the available {\it Herschel} data for the FFs has already been presented in \citet{2016MNRAS.459.1626R}, which provides an extremely valuable comparison sample. We caution that our photometry differs from that presented in \citet{2016MNRAS.459.1626R}, mainly due to the different approaches taken. A critical point here is that among the twelve high-significance ALMA-FF detections, only five have counterparts which were previously identified in the {\it Herschel} study. Thus while we fit the SPIRE emission assuming a principal association with the detected ALMA sources, \citet{2016MNRAS.459.1626R} typically identify bright galaxy counterparts based on IRAC, MIPS 24\,$\mu$m and PACS 100\,$\mu$m locations and assume all of the SPIRE emission is associated with these. As the majority of the \citet{2016MNRAS.459.1626R} identifications have $z$$\lesssim$1, their FIR SEDs should peak at 100--200\,$\mu$m and contribute only weakly beyond 350\,$\mu$m.  The ALMA 1.1\,mm sources, by contrast, are likely to lie at $z$$\gtrsim$1 and contribute substantially at $\sim$250--500\,$\mu$m; thus they are critical to use as priors for deblending the SPIRE emission. Without accounting for differences in the fitting methods themselves, the main differences come from the number of sources used for the deblending and the identification of such sources.

%

\begin{figure*}
\centering
\includegraphics[angle=270, width=16cm]{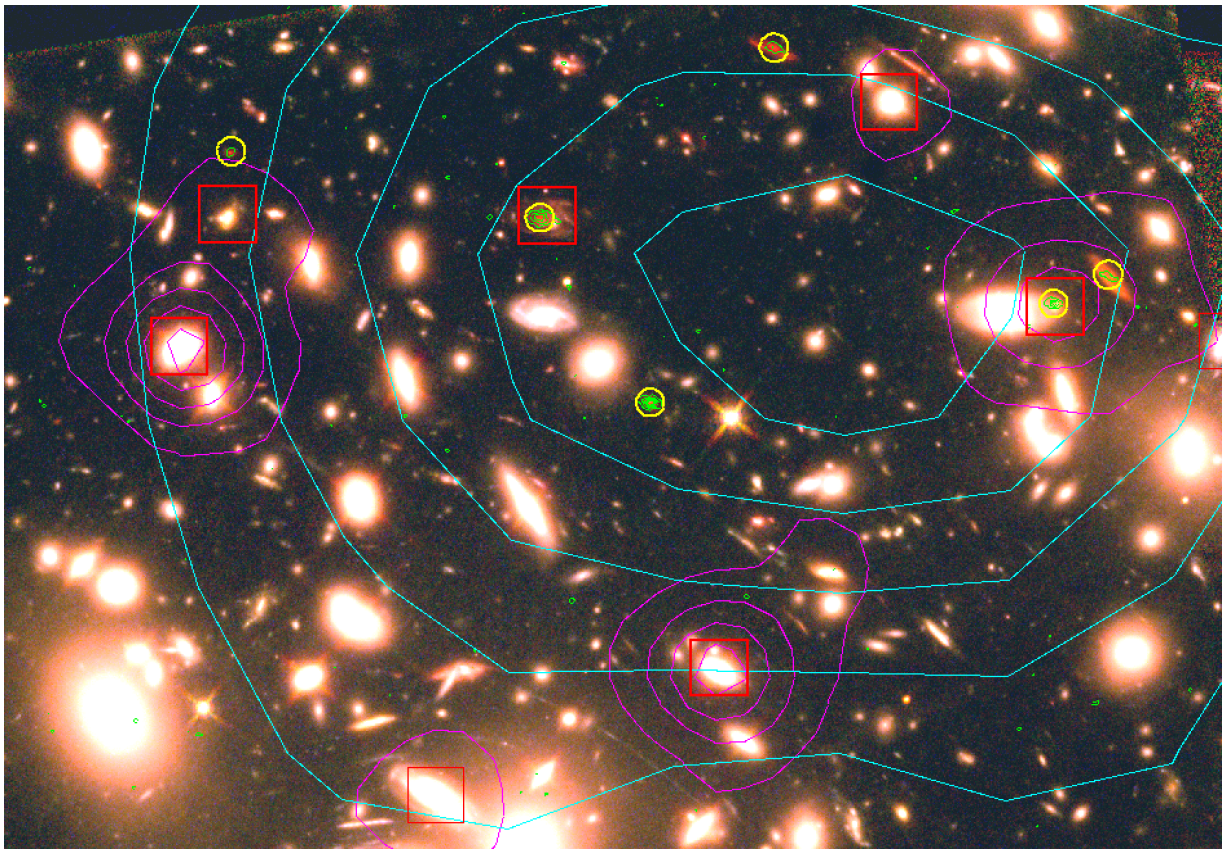}
\caption{\label{rawle1} {\it HST} color image (F814W as blue, F125W as green, and F160W as red) showing the NW corner of the galaxy cluster A2744. The red squares denote the FIR counterparts found by \citet{2016MNRAS.459.1626R}, while the yellow circles correspond to the ALMA detected sources. The green, cyan, and magenta contours show emission in the ALMA 1.1\,mm, SPIRE 500\,$\mu$m, and PACS 100\,$\mu$m images, respectively. The low overlap ($\sim$40\%) between the PACS and ALMA samples highlights the potential difficulty and disconnect of making identifications based on mid-IR priors combined with low resolution and blended {\it Herschel} imaging alone. The resulting deblended photometry can be very different when longer wavelength counterparts are incorporated into the process. In particular, a large fraction of the SPIRE 500\,$\mu$m emission shown here arises from the ALMA-detected counterparts and not from the PACS-detected ones.\label{fig:rawle1}. As in Fig. 1, the ALMA 1.1mm green contours start at 4$\sigma$ (220 $\mu$Jy/beam) and increase in 2 $\sigma$ increments (110 $\mu$Jy/beam). The SPIRE 500$\mu$m cyan contours start at 0.005 Jy/beam and increment by 0.0034 Jy/beam. And the PACS 100$\mu$m magenta contours start 0.08 Jy/beam and increment by 0.12 Jy/beam.}
\end{figure*}
        
\begin{figure*}
\centering
\includegraphics[angle=270, width=16cm]{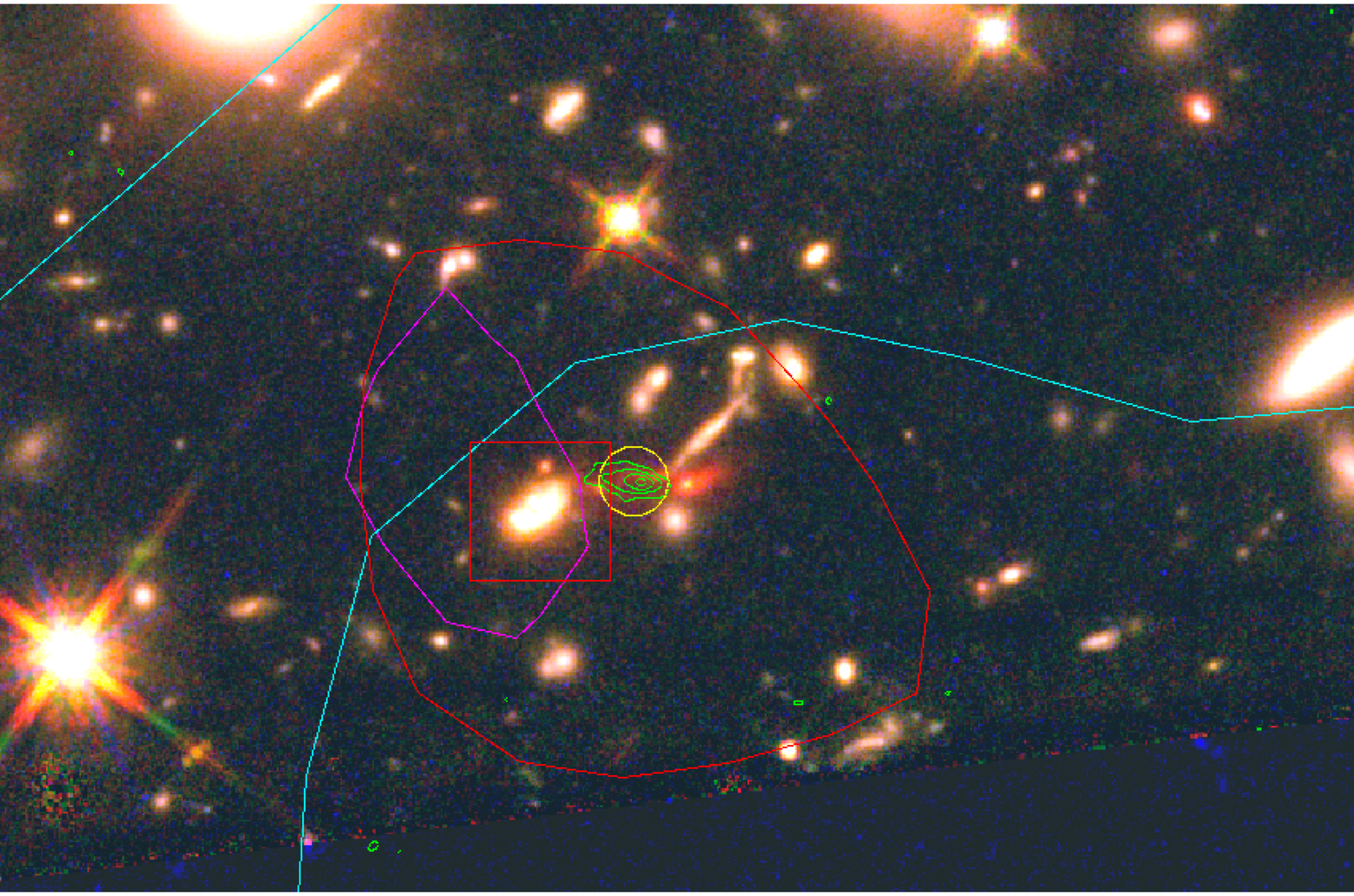}
\caption{\label{rawle2} {\it HST} color image showing the vicinity of A2744-ID02, highlighting the potential for misidentification of FIR counterparts even for relatively isolated objects. Symbols and contours same as Fig.~\ref{fig:rawle1}, with the addition of red contours denoting emission at 160\,$\mu$m. {\bf PACS 160$\mu$m contours start at 0.08 Jy/beam and increments by 0.16 Jy/beam.} While the 100\,$\mu$m contours are offset from A2744-ID02, the 160\,$\mu$m (and all longer wavelength) emission is centered on the ALMA source, indicating that most of the FIR emission assigned to the galaxy denoted by the red square in fact is likely to be associated with the ALMA-detected galaxy. \label{fig:rawle2}}
\end{figure*}
        
The differences in the FIR fitting methodology are most significant in cluster A2744, which has the highest number of ALMA-detected sources. Figure~\ref{fig:rawle1} compares the spatial distributions of the \citet{2016MNRAS.459.1626R} counterparts and ALMA-detected sources. The SPIRE 500\,$\mu$m contours (cyan) are well-aligned with the ALMA 1.1\,mm sources. The shift in prior positions results in very different deblended SPIRE flux densities for the associated galaxies, with a number of the \citet{2016MNRAS.459.1626R} sources changing from strong SPIRE detections to upper limits. In Fig.~\ref{rawle2}, we show a second example, this time of a relatively isolated {\it Herschel} source from \citet{2016MNRAS.459.1626R} in A2744. The FIR emission had been assigned to the optically bright galaxy denoted by the red square to the left, while the optically faint, red galaxy beneath the green ALMA contours to the right corresponds to ALMA source A2744-ID02 (one of the brightest 1.1\,mm sources detected in the A2744 mosaic). Given the estimated redshift (see $\S$\ref{section.sed}), it is very probable that the vast majority of the FIR emission in Fig.~\ref{rawle2} is associated with A2744-ID02 rather than the \citet{2016MNRAS.459.1626R} counterpart. These cases demonstrate how critical long-wavelength observations can be for the deblending of complex SPIRE emission. Ideally, higher frequency ALMA observations, well-matched to the SPIRE bands, would allow us to remove fully the remaining modeling degeneracies and more firmly establish the fraction of emission associated with the ALMA and PACS counterparts, respectively.

\section{Physical properties}
\label{section.sed}
In the following section, we estimate photometric redshifts for the ALMA detected sources based on SED-fitting. We independently consider two sets of photometry, in order to investigate degeneracies that arise when only a portion of the SED is assessed:
\begin{itemize}
\item  \textit{HST} $F435W$ to IRAC 8.0\,$\mu$m bands (NIR-SED) 
\item  MIPS 24\,$\mu$m to ALMA 1.1\,mm (FIR-SED)
\end{itemize}
We then attempt to fit complete SEDs from 0.4\,$\mu$m to 1.1\,mm, and conclude with a discussion of the most reliable photometric redshifts and basic properties of the ALMA-FF DSFGs. 

\subsection{Photometric redshifts}

As already demonstrated by several authors (e.g. \citealt{2011A&A...534A.124B}) the estimated physical properties, and more especially the photometric redshifts, of sources detected at sub-mm wavelengths can strongly differ according to the wavelength range considered in the SED-fitting analysis. The ALMA-FF DSFGs, however, are probing an order of magnitude fainter in flux, and thus it is useful to understand how this affects such biases.

We used an updated version of \textit{Hyperz} \citep{2000A&A...363..476B} to estimate the photometric redshifts of these sources. We define the parameter space as follows. We consider a redshift range from 0.0--6.0, since all our objects are detected on the \textit{HST} $F814W$ and/or $F105W$ images. The reddening interval is defined between $A_v$=0.0--20\,mag, considering that some ALMA-detected sources could be strongly affected by dust (e.g., A2744-ID02). We adopt a template-based method using template library models from \citet{1997A&A...326..950F}, \citet{1998ApJ...509..103S}, \citet{2003MNRAS.344.1000B}, \citet{1980ApJS...43..393C}, \citet{1996ApJ...467...38K}, \citet{2007ApJ...663...81P}, \citet{2010A&A...514A..67M} and \citet{2001ApJ...556..562C}. We also added to our library ULIRG templates published in \citet{2008A&A...484..631V}, and three templates built from the ALESS survey \citep{2015ApJ...806..110D}.

We first derive photometric redshifts using only the wavelength range $\approx$0.4--8\,$\mu$m. The majority of our objects display a best SED-fit at $z$$\leq$3, and only one detected source, namely A2744-ID04 prefers a $z\gtrsim$4 solution. Interestingly, none of the ALMA-detected sources has a best NIR-SED-fit corresponding to one of the ULIRGs templates in our library. 

We then estimated photometric redshifts of the ALMA-FFs sources using only the FIR-SEDs with templates covering the FIR peak \citep[e.g.,][]{2007ApJ...663...81P}. A caveat here is that most of the sources have few robust constraints in the FIR, making it more difficult to estimate FIR photometric redshifts. Therefore, we considered two different approaches to fit the FIR-SEDs: 
\begin{itemize}
\item using the true upper limits as measured from the MIPS, PACS, and SPIRE data. In that case, the flux in these bands are set to $F_{obs}=0$ with an error bar equal to the limiting flux;
\item setting the flux in each FIR band to half of the 3$\sigma$ upper limits and the error bar set to 50\% of the 3$\sigma$ limit, considering that each ALMA source is the main contributor to the FIR flux and the true flux lies just below the detection limit ("pseudo-detections"). By forcing the limits to be detections, we are assigning a different weighting to the error distribution. Given that adjacent bands have detections, we expect the true flux to be closer to the upper limit than to zero.
\end{itemize}
Four objects could not be fitted properly due the small number of constraints in the FIR (namely A2744-ID2, A2744-ID3, A2744-ID05 and M0416-ID04) and the estimates for these sources are more uncertain. All the remaining objects are well fitted at $z$$\sim$2--3 with no object above $z$$\sim$3.5. Table~\ref{tab.photz} provides results for all sources. We note that based on the reduced $\chi^2_{\nu}$, fits using the true upper limits (except for A2744-ID07 and M1149-ID01) are better.

Finally we used the combined NIR$+$FIR SED constraints to estimate photometric redshifts, again considering FIR constraints in two ways: true upper limits and pseudo-detections as described above. For about 85\% of our sample, the photometric redshift estimates are consistent with those found for the FIR-SEDs alone, but with higher reduced $\chi^2_{\nu}$. For the remaining objects, the lack of good photometric templates spanning the full range of wavelength explored in this study leads to worse fits than with the NIR-SEDs or FIR-SEDs alone. 

By combining the results obtained using the three different SED-fitting trials (NIR-SED, FIR-SED and combined NIR+FIR-SED), and assuming that the FIR shape provides strong clues on the true photometric redshift of our sources \citep[e.g.,][]{2011A&A...534A.124B, 2011A&A...531A..74L}, we obtain reasonable estimates on the photo-$z$ for all our sample. We applied the following procedure in order to estimate the best photometric redshift of our targets : 
\begin{itemize}
\item Estimate photo-$z$ from NIR-SED, FIR-SED, and FULL-SED
\item If reduced $\chi^2_{\nu}$ of NIR-SED and FULL-SED are $>$ 2, then adopt FULL-SED, which may have worse $\chi^2_{\nu}$ but at least has very wide range of possible values likely to encompass real redshift.
\item If $\chi^2_{\nu}<$2, and NIR-SED and FULL-SED provide consistent ranges, adopt one with lowest $\chi^2_{\nu}$
\item If $\chi^2_{\nu}<$2 but NIR-SED and FULL-SED are not consistent within ranges, adopt one that is consistent with FIR-SED.
\item If $\chi^2_{\nu}<2$ but no estimates are consistent, adopt FIR-SED, which may have worse $\chi^2_{\nu}$ but at least has very wide range of possible values likely to encompass real redshift.
\end{itemize}
Two objects among our sample have a FIR-SED not well fitted (reduced $\chi^2_{\nu} >$2.0), namely A2744-ID06 and M0416-ID01. For these two objects, we applied the previous procedure but we keep in mind that the deduced properties for A2744-ID06 are subject to caution (M0416-ID01 has a secure spectroscopic redshift, see below) . 

The combined results are listed in Table~\ref{tab.photz} and are used to estimate further properties of the ALMA sources listed in Table~\ref{tab.photometry}. 


As a crosscheck on the reliability of our photometric redshift estimates, we used another SED-fitting tool, \textit{iSEDfit} \citep{2013ApJ...767...50M}, on the NIR-SED. We find general consistency between the \textit{Hyperz} and \textit{iSEDfit} results. 

\subsection{Spectroscopic Redshifts}
\label{sec.spectro}

All of the \textit{FFs} clusters have been observed within the framework of the \textit{Grism Lens-Amplified Survey from Space} project (hereafter \textit{GLASS} - Cycle 21 ID : 13459, PI: T. Treu) combining three \textit{HST} grisms: G800L, G102 and G141 (\citealt{2015ApJ...812..114T}, \citealt{2014ApJ...782L..36S}). We compared the redshift catalogs released by the team for A2744, M0416, and M1149 to our photometric redshifts. Among all the ALMA-FFs DSFGs presented here, only five objects have \textit{GLASS} redshifts:

\begin{itemize}
\item A2744-ID01: the best photometric redshift is $z_{\rm phot}$=2.95$^{+0.45}_{-1.82}$ while \textit{GLASS} obtains a low-quality spectroscopic redshift of $z_{\rm spec}$$\sim$2.9 based on a "red continuum" feature. These are in relatively good agreement.
\item A2744-ID02: the best photometric redshift is $z_{\rm phot}$=2.02$^{+0.18}_{-0.88}$, while \textit{GLASS} obtains a good-quality spectroscopic redshift of $z$=2.482 based on the detection of the 4000\AA\ break. The \textit{GLASS} redshift is inconsistent with our photometric estimate from the NIR-SED (Fig.~\ref{fig.SED}).
\item M0416-ID01: the best photometric redshift is $z_{\rm phot}$=2.23$^{+0.07}_{-0.03}$ based on the NIR$+$FIR SED, while \textit{GLASS} obtains a high-quality spectroscopic redshift of $z$=2.086 based on the detection of the [O{\sc iii}], H$\beta$ and [Mg{\sc ii}] emission lines. The \textit{GLASS} spectrum classifies this object as an AGN and is consistent with the photometric redshifts deduced from both the NIR-SED and NIR$+$FIR SED fits (Fig.~\ref{fig.SED}). 
\item M0416-ID02: the best photometric redshift is $z_{\rm phot}$=2.13$^{+0.33}_{-0.36}$, while \textit{GLASS} obtains a good-quality spectroscopic redshift of $z_{spec}$$=$1.953 based on the detection of the 4000\AA\ break. These are in relatively good agreement.
\item M1149-ID01: the best photometric redshift is $z_{\rm phot}$=1.12$^{+1.18}_{-0.61}$, while \textit{GLASS} obtains a good-quality spectroscopic redshift of $z_{spec}$$=$1.460 based on the detection of the 4000\AA\ break. The \textit{GLASS} redshift is relatively consistent with the photometric redshift we estimated from all the photometric constraints we used in this study.
\end{itemize}
\noindent Based on the above, we adopt \textit{GLASS} redshifts where available and use our best $z_{\rm phot}$ estimates otherwise.

Figure~\ref{fig.histo_z} compares the redshift distribution of the ALMA-FFs DSFGs with those from several previous ALMA-selected samples \citep{2015ApJ...806..110D, 2016arXiv160600227D, 2016arXiv160706769A}. The ALMA-FFs sample span a moderate redshift range between 1.0--2.9, with a mean photometric redshift of <$z_{\rm phot}$>=1.99$\pm$0.27, where the error bar reflects the standard deviation. Given the small number statistics, this value is consistent with what has been found in deep ALMA imaging of the \textit{Hubble} Ultra Deep Field (HUDF) reaching a 5$\sigma$ limit of 175 $\mu$Jy \citep[<$z$>=2.1$\pm$1.0;][]{2016arXiv160600227D} and the extremely deep ALMA Spectroscopic Survey in the HUDF (ASPECS) reaching a 5$\sigma$ limit of 65 $\mu$Jy  \citep[e.g., <$z$>=1.6$\pm$0.4;][]{2016arXiv160706769A}. Such an evolution in the mean redshift distribution of DSFGs as a function of the flux density has been predicted by \citet{2015A&A...576L...9B} using a phenomenological model. Based on this model, at the observed 5$\sigma$ limit of $\sim$0.28-0.35 mJy for the 1.1mm sources in the ALMA-FFs, the expected mean redshift distribution of our survey is $<z_{phot}>\sim$2.0, which is fully consistent with our result.

\begin{figure}
\centering
\includegraphics[width=9cm]{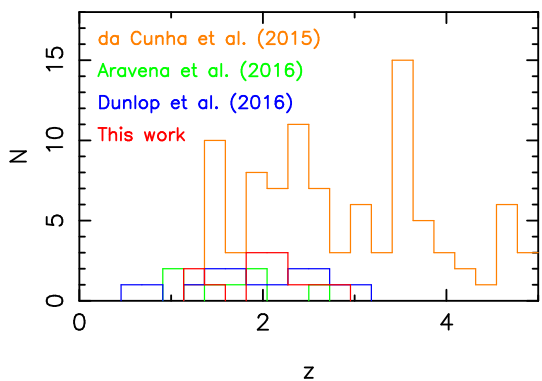}
\caption{\label{fig.histo_z} Redshift distributions (photometric or spectroscopic when available) for the ALMA-FFs sample compared with those obtained for other deep ALMA imaging surveys: the \textit{Hubble} Ultra Deep Field (grey, \citealt{2016arXiv160600227D}); the ALMA Spectroscopic sample (green, \citealt{2016arXiv160706769A}) and the ALESS survey (blue, \citealt{2015ApJ...806..110D}).}
\end{figure}



\addtolength{\tabcolsep}{-3pt}
\begin{table*}
\scriptsize
\caption{Photometric Redshift Estimates for the ALMA-FFs Sources}             
\label{tab.photz}      
\centering                          
\begin{tabular}{|c|cc|crc|crc|crc|crc|crc|}        
\hline\hline                 
 & & & \multicolumn{3}{c|}{NIR-only constraints} & \multicolumn{6}{c|}{NIR$+$FIR constraints} & \multicolumn{6}{c|}{FIR-only constraints} \\ 
 \cline{7-18} ID & RA & DEC & \multicolumn{3}{c|}{} & \multicolumn{3}{c|}{(Upper limits)} & \multicolumn{3}{c|}{(detection)} & \multicolumn{3}{c|}{(Upper limits)} & \multicolumn{3}{c|}{(detection)} \\
\cline{4-18}  & [J2000] & [J2000] & z$_{\rm phot}$ & $\chi^2_{\nu}$  & 1$\sigma$ & z$_{\rm phot}$ & $\chi^{2}_{\nu}$  & 1$\sigma$ & z$_{\rm phot}$ & $\chi^{2}_{\nu}$  & 1$\sigma$
& z$_{\rm phot}$ & $\chi^{2}_{\nu}$  & 1$\sigma$ & z$_{\rm phot}$ & $\chi^{2}_{\nu}$  & 1$\sigma$ \\
 \hline                        

A2744-ID01	& 00:14:19.8 & $-$30:23:07.6 & 2.95$^{\star}$ 	&	0.41	&	2.50 - 4.77	&	2.46 	&	0.64	&	2.2 - 2.3	&	3.11	&	0.66	&	2.19 - 3.34 & 3.69  &  0.64 &  2.46  -  5.00  &  1.53  &  0.54  &  0.47 - 5.74	\\
A2744-ID02	& 00:14:18.2& $-$30:24:47.3 &	1.35	&	1.94	&	1.31 - 1.41	&	2.02$^{\star}$	&	3.44		&	1.1 - 2.2.	&	2.02	&	3.39	&	1.70-2.20 & -  &  -  &  2.58  -  8.00  &  3.11  &  0.05  &  0.37 - 6.0	\\
A2744-ID03	& 00:14:20.4 & $-$30:22:54.6 & 2.52$^{\star}$	&	0.72	&	2.07 - 2.75	&	2.45	&	1.09	&	2.32	-	2.53	&	2.43	&	0.86	&	1.76 - 2.59  & -     &   -  &  2.20 - 8.00   &  1.23  &  0.20  &  0.24 - 6.00	\\
A2744-ID04	& 00:14:17.6 & $-$30:23:00.7 & 1.21	&	3.81	&	0.93 - 1.44	&	0.79	&	2.12	&	0.60 - 1.00	&	0.79	&	2.12		&	0.60 - 0.90  & 1.44  &  4.81  &  1.32  -  1.55  &  1.02$^{\star}$  &  1.01  &  0.93 -1.34	\\
A2744-ID05	& 00:14:19.1 & $-$30:22:42.2 & 2.01$^{\star}$	&	0.28		&	1.85 - 2.70	&	2.34	&	1.74	&	1.93 - 2.44	&	1.72	&	0.64	&	0.70 - 2.04  & -  &  -  &  2.08 -  8.0   &  1.24  &  0.30  &  0.16 - 6.00 	\\
A2744-ID06	& 00:14:17.3 & $-$30:22:58.7 & 2.08$^{\star}$	&	1.33	&	2.00 - 2.21	&	1.30	&	3.58	&	1.20 - 1.40	&	1.30	&	3.57	&	1.20 - 1.40 	& 2.24  &  8.81  &  1.98  -  2.51  &  2.27  &  7.63  &  1.54 - 2.63 \\
A2744-ID07	& 00:14:22.1 & $-$30:22:49.8 & 2.95	&	0.69	&	2.93-3.04	&	1.85$^{\star}$	&	0.41	&	1.71 - 2.01	&	1.70	&	0.54	&	1.36 - 2.60	&  2.14    &  2.45     &  1.66  -  8.0   &  1.98  &  1.55  &  0.70 - 2.39 \\
M0416-ID01	& 04:16:10.8 & $-$24:04:47.5 & 2.23$^{\star}$	&	1.68	&	2.20-2.30	&	1.17	&	2.64	&	1.00 - 1.90	&	1.70	&	2.76	&	1.40 - 1.91 &  1.40  &  0.51  &  1.09  -  1.73  &  1.93  &  5.72  &  1.22  -  2.36 \\
M0416-ID02	& 04:16:07.0 & $-$24:03:59.9 & 2.13$^{\star}$	&	1.17	&	1.77-2.46	&	1.29	&	1.43	&	1.10 - 1.47	&	1.30 &	1.55	&	1.10	-	1.50 	&  1.64  &  0.43  &  1.23  -  2.19  &  2.03  &  1.96  &  1.53  -  2.44 \\
M0416-ID03	& 04:16:08.8 & $-$24:05:22.4 & 1.34	&	2.86	&	1.13 - 1.44	&	1.29$^{\star}$	&	1.36	&	0.90 - 1.40	&	1.30	&	1.41	&	1.00 - 1.40	&  1.47  &  0.57  &  1.01  -  2.08  &  2.00  &  1.92  &  1.60  -  2.34 \\
M0416-ID04	& 04:16:11.7 & $-$24:04:19.6 & 2.27$^{\star}$	&	0.78	&	1.66 - 2.44	&	2.07	&	2.47	&	1.96	-	2.19	&	1.61	&	0.82	&	1.41 - 2.08 	&  -  &  -  &  1.26  -  8.00  &  2.09  &  0.29  &  1.78  -  2.39 \\
M1149-ID01	&	11:49:36.1 & $+$22:24:24.5 & 1.12$^{\star}$	&	0.24	&	0.51 - 2.30	&	1.24	&	1.04	&	1.00 - 1.50	&	1.22	&	1.11	&	1.00 - 1.40,.	&  1.92  &  0.27  &  0.58  -  2.36  &  1.86  &  0.17  &  1.37  -  2.48 \\
\hline
\end{tabular}
\begin{flushleft}
Columns: (1) ID; (2,3) RA,Dec ; (4) Photometric redshift, (5) $\chi^2_{\nu}$, and (6) 1$\sigma$ confidence interval, respectively, associated with the best fit of the NIR-SED only; (7,8,9) same parameters for the best fit of the full SED (NIR-SED$+$FIR-SED) assuming upper limits for the MIPS, PACS, and SPIRE photometry; (10, 11, 12) same parameters for the best fit of the full SED (NIR-SED$+$FIR-SED) but forcing a detection in the MIPS, PACS, and SPIRE bands (see text for details); (13, 14, 15) same parameters for the best fit of the FIR-SED only, assuming upper limits for the MIPS, PACS, and SPIRE photometry; (16, 17, 18) same parameters for the best fit of the FIR-SED only but forcing a detection in the MIPS, PACS, and SPIRE bands (see text for details). \\
$^*$ Preferred solution (see text for details)
\end{flushleft}
\end{table*}
\addtolength{\tabcolsep}{3pt}

\subsection{Colors}

\begin{figure*}
\centering
\hglue-0.5cm{\includegraphics[width=10.5cm]{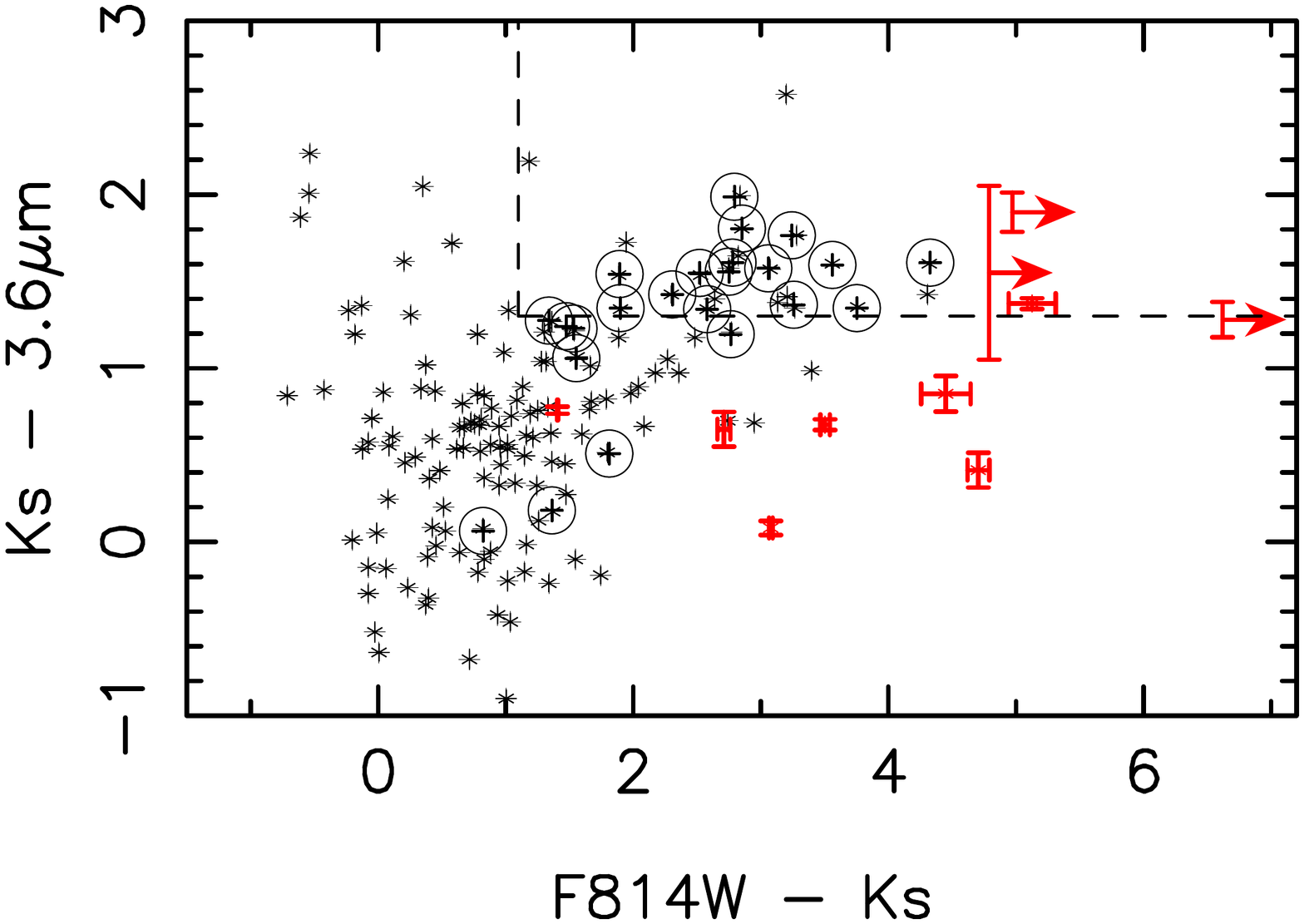}}
\hglue-2.2cm{\includegraphics[width=10.5cm]{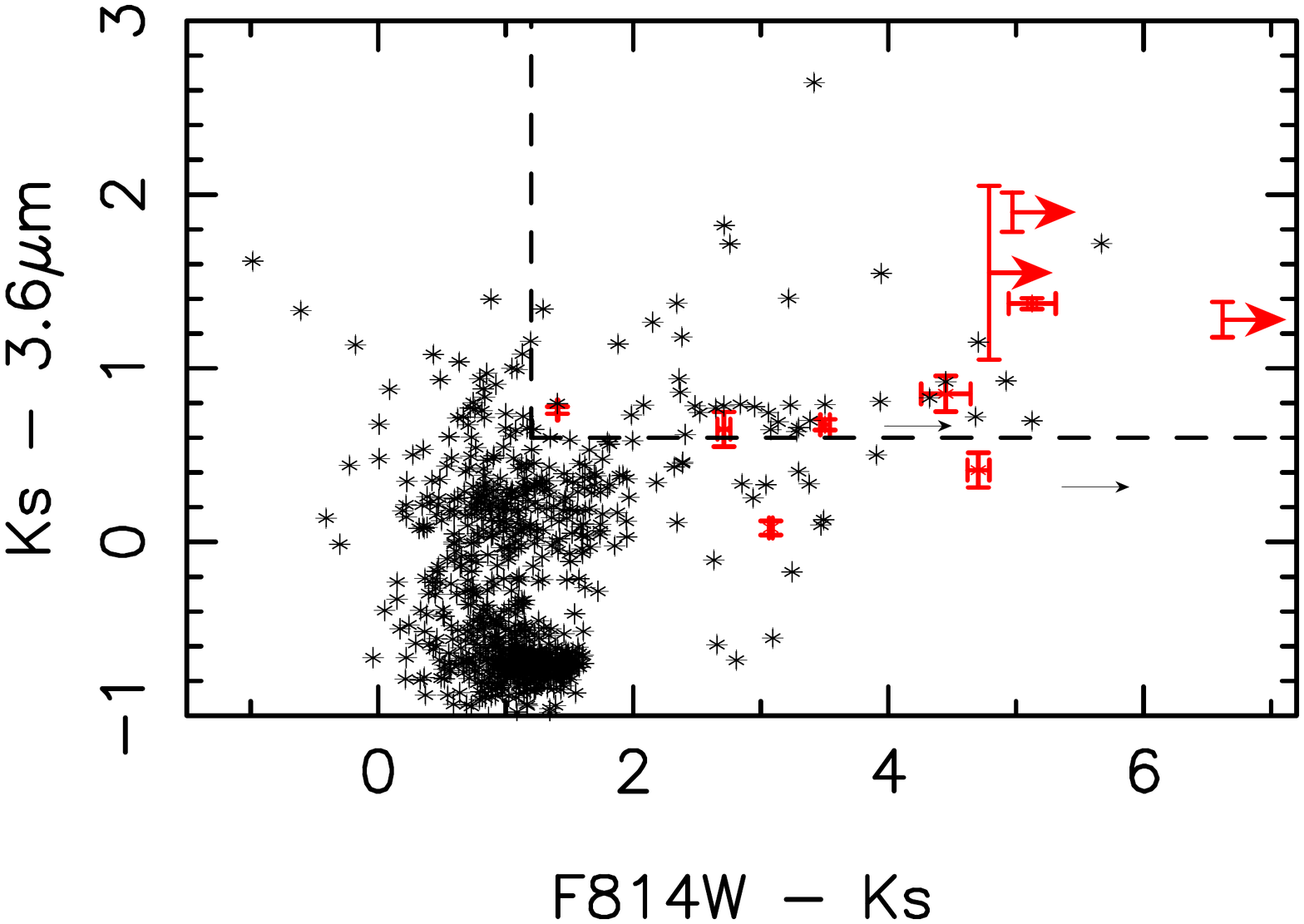}}
\caption{\label{fig.color} $F814W-K_{\rm s}$ vs. $K_{\rm s}-[3.6]$ colors for the ALMA-FFs DSFGs (red symbols, both panels). 
On the {\it left}, we compare to 22 ALMA-detected submm galaxies (SMGs, black circles + crosses) and 142 non-SMG field sources (black crosses) presented by \citet[][C16]{2016ApJ...820...82C}. The nominal OIRTC cuts adopted by C16 are denoted by the dashed region. The ALMA-FFs DSFGs appear systematically offset from the C16 sample by 0.6--1.0 mag in $K_{\rm s}-[3.6]$, likely due to different aperture correction choices.
On the {\it right}, we compare 11 ALMA-FFs DSFGs to 767 non-stellar $>$5$\sigma$ sources from the \textit{AstroDeep} $F160W$ catalog (black crosses) in the ALMA footprints of the A2744 and M0416 clusters. Following the prescription in C16, we recalibrate the OIRTC color cuts for the \textit{AstroDeep} sample, which are again denoted by the dashed region. In both figures, the 11 objects with detections in $K_s$ are plotted; two objects, namely A2744-ID04 and M0416-ID04, have similar colors and therefore lie on top of each other in these figures.}
\end{figure*}

As mentioned in $\S$\ref{sec.counterparts}, the majority of the counterparts are quite red. This is not surprising, given that previous studies have already demonstrated that DSFGs typically have red optical-NIR colors in $i - K$, $J - K$, $K - [4.5]$, etc., foreshadowing their high-$z$, dusty nature \citep[e.g.,][]{2002MNRAS.331..495S, 2004ApJ...606..664D}. The sample of DSFGs detected in the ALMA-FFs clearly follows these trends, as shown in Fig.~\ref{fig.color}. We note that eight of the 11 objects with $Ks$ measurements have $F814W-K_{s}\gtrsim 4$ (the classic extremely red object cut) and four have $F160W-[4.5]\gtrsim3$ .

\citet[][hereafter C16]{2016ApJ...820...82C} recently investigated the use of color selection to isolate DSFGs, settling on the Optical–IR Triple Color (OIRTC) method using $z-K$, $K-[3.6]$, and $[3.6]-[4.5]$ colors to select most optimally submm/mm detected counterparts. The accuracy ($N_{\rm confirmed}$/$N_{\rm selected}$$\sim$90\%) and completeness ($N_{\rm confirmed}/N_{\rm total}$$\sim$50\%) of the dynamic OIRTC method in the C16 study rivals that of the more traditional radio identification \citep[e.g.,][]{1999ApJ...513L...9R}, but beneficially can select fainter near-IR sources. Intriguingly, only three of the 11 ALMA-detected galaxies in the FFs satisfy the nominal OIRTC color cuts found by C16, with nearly all of the remaining objects failing the $K_{\rm s}-[3.6]>1.25$ cut in particular (Fig. ~\ref{fig.color}, {\it left} panel). Additionally, half of the ALMA-FFs DSFGs appear redder in $F814W-K_{\rm s}$ than the reddest sources in C16, some by at least $\sim$1--2 mag.

To understand these differences better, we also plot the colors of 767 sources within the ALMA-FFs footprints of clusters A2744 and M0416 from the \textit{AstroDeep} catalogs, selected to have $>$5$\sigma$ confidence detections in the $F160$, $K_{\rm s}$, and $[3.6]$ images and stellarity parameters of $<$0.9 (Fig. ~\ref{fig.color}, {\it right} panel).\footnote{There are roughly an equivalent number of additional FFs galaxies with limits in $F160$, $K_{\rm s}$, and/or $[3.6]$. However, nearly all of these are uninteresting in the sense that they are highly unlikely to occupy the region of color space where DSFGs lie. For visualization purposes and simplicity, we exclude these from discussion.} Due to the 6.6$\times$ larger survey area and 1.7 mag deeper K-band imaging compared to C16, our field galaxy sample is 5--10 times larger and includes a large number of cluster galaxies (clumped around $F814W-K_{\rm s} \sim 1.2$ and $K_{\rm s}-[3.6]\sim -0.8$). Regardless of an ACS (e.g., $F814W$) detection, only eight field galaxies fall in the original selection box defined by C16, again largely due to the $K_{\rm s}-[3.6]$ cut. As mentioned above, only three of these are ALMA-detected, implying low selection accuracy (38\%) and completeness (27\%). However, the full \textit{AstroDeep} sample appears displaced in $K_{\rm s}-[3.6]$ by $\sim$0.6--1.0 mag from that of C16, likely due to the different methods adopted for performing {\it Spitzer} IRAC aperture corrections (C.-C. Chen, private communication). Thus a degree of caution must be exercised when employing color cut selection blindly. To compensate for this, we recalibrated the OIRTC method of C16 directly based on the \textit{AstroDeep} photometry, arriving at new color cuts of $F814W-K_{\rm s}>1.2$, $K_{\rm s}-[3.6]>0.6$, and $[3.6]-[4.5]>0.16$. Based on these new cuts, 10 ALMA-detected DSFGs are selected among 28 total field galaxies, yielding a formal accuracy (30\%) and completeness (80\%). An additional seven objects selected by the new OIRTC cuts are detected at lower significance ($\gtrsim$4$\sigma$) in the ALMA maps and another lies within 2\arcsec of A2744-ID02, implying a higher accuracy ($\sim$65\%) at similar completeness, more in line with the values found by C16. For the remaining ten objects selected by the OIRTC cuts,
it is not immediately clear why they lack significant ALMA emission, although they do show a $S/N$$\sim$3.8 when stacked in aggregate (R. Carvajal et al., in preparation). Given the similarity in stellar properties to the ALMA detections, they likely represent the less dust-obscured tail of the parent distribution.

In conclusion, we confirm that the C16 OIRTC method appears to provide a very efficient way of pre-selecting ALMA-detected DSFGs.


\subsection{Magnification}

One of the main advantages of the FFs survey is that several teams have provided amplification maps for all the clusters, following different assumptions to produce mass models \citep{2011MNRAS.417..333M,2014MNRAS.444..268R,2014ApJ...797...48J,2015ApJ...800...38G,2015ApJ...801...44Z}. We used \textit{Lenstool} \citep{2009MNRAS.395.1319J} to calculate magnifications for the ALMA DSFGs, adopting the parameters files defined by the CATS (\textit{Cluster As TelescopeS}) group, which has incorporated the detection of several hundred multiple images from the ACS and WFC3 FFs datasets \citep[e.g.,][]{2014MNRAS.443.1549J,2015MNRAS.452.1437J}. 
The choice of the CATS models is supported by their overall performances as discussed in \citet{2016arXiv160604548M}. We used the photometric redshifts deduced from SED-fitting to estimate the magnifications of these objects, except for objects with robust spectroscopic redshifts (namely A2744-ID02, M0416-ID01, M0416-ID02 and M1149-ID01). All these sources have moderate amplification factors ($\mu$), with only one object at $\mu>4$. All amplification factors are listed in Table~\ref{tab.properties}. Because boost from gravitational lensing is relatively mild for all of our sources, flux determinations and related parameter estimates for the FFs DSFGs should be relatively robust (e.g., we are not missing extended flux from the galaxies). We also used \textit{Lenstool} to check for the existence of multiple-imaged candidates, but none were predicted.

Thanks to the mean magnification factor of our sources ($\overline{\mu}$$=$2.16 $\pm$0.44), the ALMA-FFs survey effectively explores the same luminosity range as the HUDF band 6 survey of \citet{2016arXiv160600227D}. For instance, the 5$\sigma$ depth of the HUDF is 0.175 mJy, whereas the 5$\sigma$ depth of our survey ranges between (0.28--0.35)/$\mu$ mJy. However, the effective area of our survey, currently 13.8 arcmin$^{2}$/$\mu$ (ultimately 23 arcmin$^{2}$/$\mu$ through cycle 3), is 1.4 (2.4) times larger than that of the HUDF (4.5 arcmin$^{2}$).


\subsection{Star Formation Rate, Stellar mass, Dust properties and Size}

With photometric and GLASS spectroscopic redshifts in hand (see section \ref{sec.spectro} for details), we can estimate some physical properties of our objects such as their SFRs, stellar and dust masses, reddening and dust temperatures. We use the latest version of MAGPHYS \citep{2008MNRAS.388.1595D}, adapted to fit the SED of high-$z$ sources \citep{2015ApJ...806..110D}. For the five objects with spectroscopic confirmation, the redshifts are fixed and we only propagate the magnification errors, while for the remaining objects we estimate 1$\sigma$ errors from the likelihood distribution of each parameter including both redshift and magnification uncertainties. Several derived properties are shown on Table~\ref{tab.properties}.

\begin{figure}
\centering
\label{fig.M_star_histo}
\includegraphics[width=8.8cm]{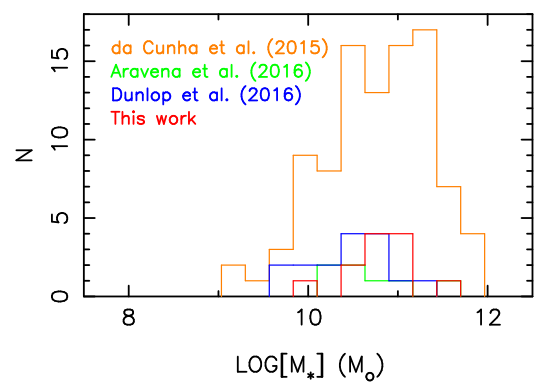}
\caption{\label{fig.M_star_histo} Distribution of the stellar masses covered by the ALMA Frontier Fields survey (red) compared with previous ALMA surveys \citep{2015ApJ...806..110D, 2016arXiv160706769A, 2016arXiv160600227D}.}
\end{figure}

Figure~\ref{fig.M_star_histo} compares the distribution of stellar masses estimated for the ALMA-FFs sample with those estimated for other recent ALMA surveys \citep{2015ApJ...806..110D, 2016arXiv160706769A}.
Consistent with previous ALMA-detected sources, the ALMA-FFs are probing the same massive population of DSFGs, with stellar masses ranging from 10.0$<\log[M_{\star}/M^{\odot}]<$11.5, and relatively high SFRs (0.9$<\log[SFR/(M_{\odot} yr^{-1})]<$2.0).   

Typical star-forming galaxies have been shown to form a tight linear relationship in the SFR-$M_{\star}$ plane out to $z$$\sim$3 \citep[e.g.,][]{2004MNRAS.351.1151B, 2007ApJ...670..156D, 2007ApJ...660L..43N, 2015ApJ...801...80L, 2015ApJ...807..141P}, dubbed the "Main Sequence", while starburst galaxies are found (by definition) to lie above the relation. Based on the positions of the ALMA-FFs DSFGs on the SFR-M$_{\star}$ diagram (Fig.~\ref{fig.SFR_vs_M_interval}), most are consistent with lying on the Main Sequence. We divided our sample into three redshift intervals, in order to better isolate any evolution between $z\sim$1.0 and 2.5, as found by \citet{2014ApJ...795..104W}. For comparison, we show all objects in public \textit{AstroDeep} catalogs, as well as the results from several recent ALMA studies. Again, the ALMA-FFs generally track some of the most massive objects between $z\sim$1.0 and 2.5. Unfortunately, the number of detected sources and large uncertainties on derived parameters do not allow strong constraints on the evolution of this distribution with redshift (see Fig. \ref{fig.M_vs_z}). 

Based on the wavelength range explored by this survey, we can also constrain the dust properties of the ALMA-FFs DSFGs. Their dust masses lie in the range 8.1$<\log[M_{dust}/M^{\odot}]<$8.8, with an average value of $\approx$8.3. These are consistent with what has been observed for the ALESS sample (see Fig.~7 of \citealt{2015ApJ...806..110D}). Their dust attenuations range from $A_{\nu}$=1.2--5.2 mag, with an average value of $<A_{\nu}>$$=$2.8$\pm$0.55 mag. This is a bit higher than what has been reported in the HUDF (1.52$\pm$0.75 mag; \citealt{2016arXiv160600227D}) but is consistent with ALESS sources ($A_{\nu}$=0.3--6.6 mag, with $<A_{\nu}>$=2.3$^{+0.8}_{-1.4}$ mag; \citealt{2015ApJ...806..110D}). Finally, we find an average dust temperature of T$_{dust}$=42$\pm$8\,K (error bars are from the standard deviation) compared to T$_{dust}$$\sim$40\,K for the ALESS sample.


\begin{figure}
\centering
\vspace{-1.0cm}
\hglue-0.5cm{\includegraphics[width=11cm]{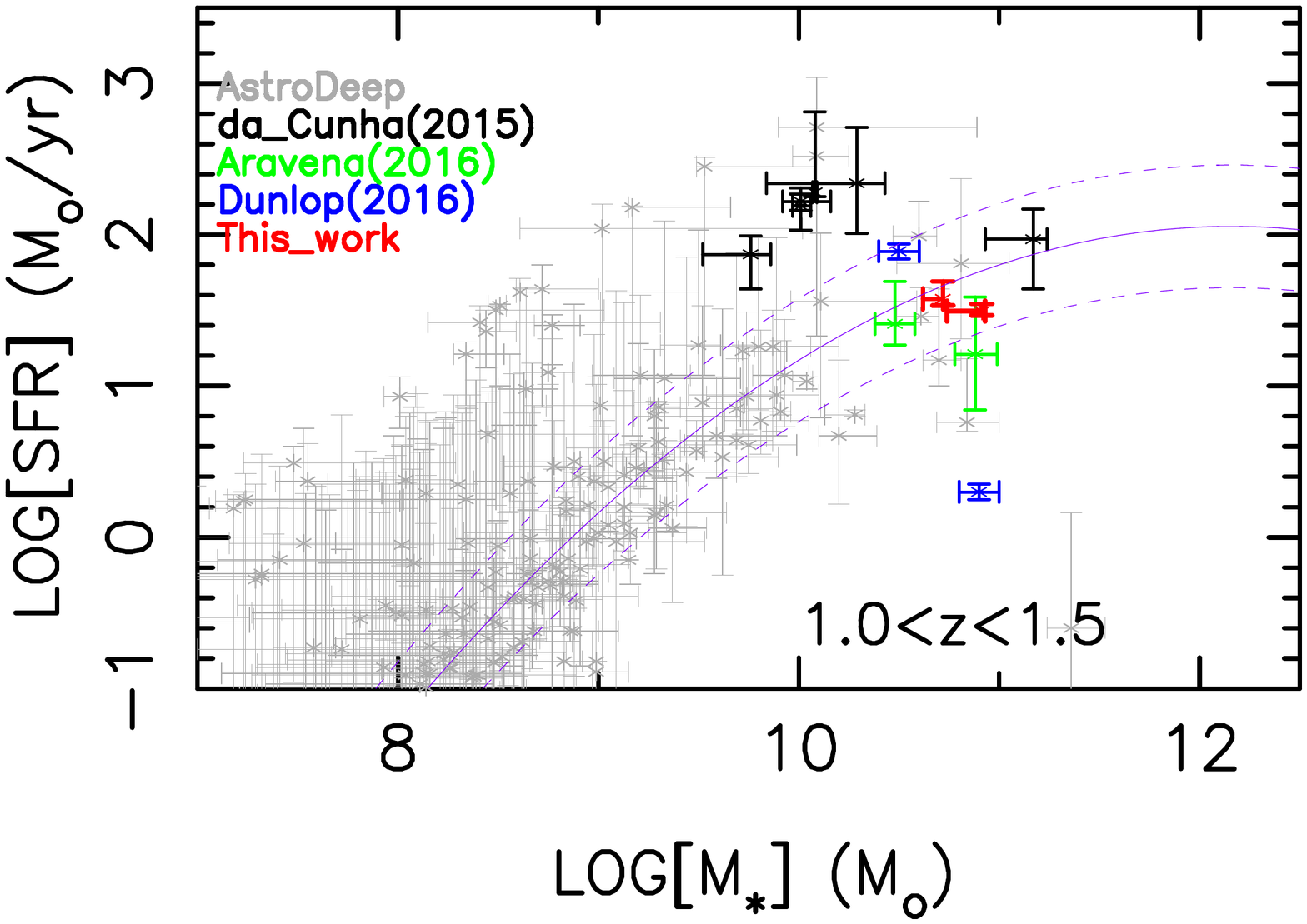}}
\vglue-2.0cm{\hglue-0.5cm{\includegraphics[width=11cm]{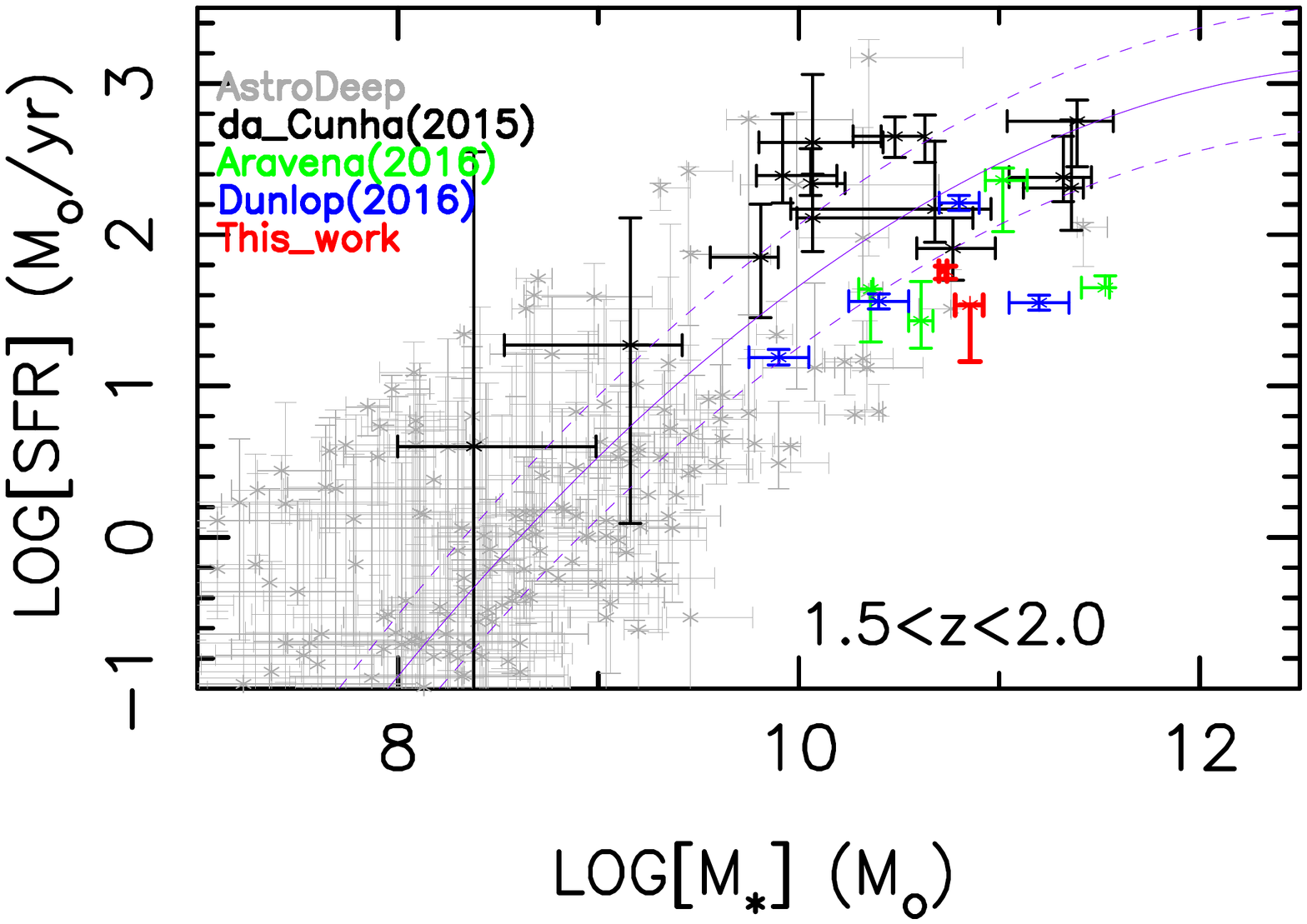}}}
\vglue-2.0cm{\hglue-0.5cm{\includegraphics[width=11cm]{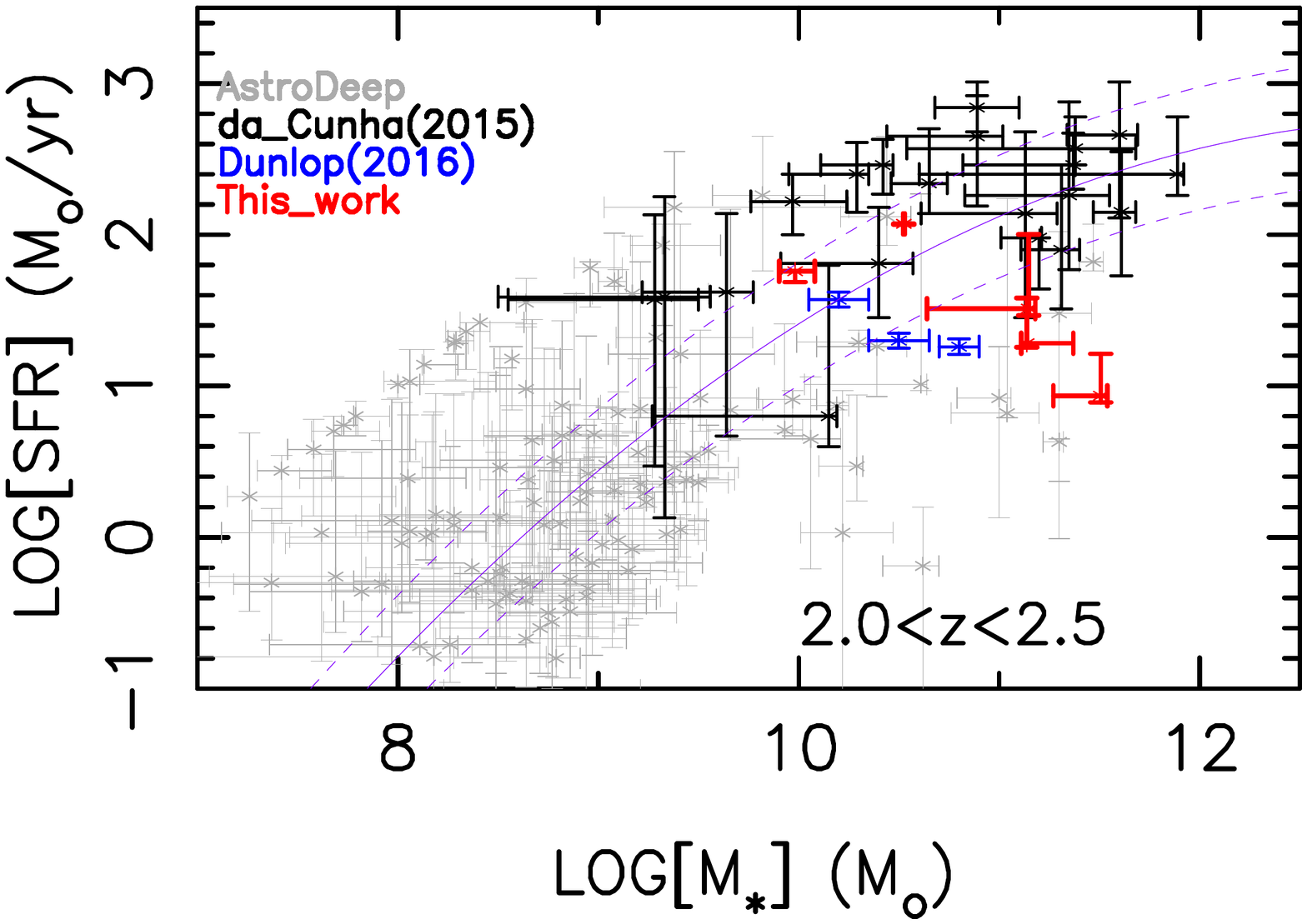}}}
\caption{\label{fig.SFR_vs_M_interval} $M^{*}$ vs. SFR for the ALMA-FFs DSFG sample (red points) and all field galaxies in the \textit{AstroDeep} FFs catalogs (grey points), divided into three redshift ranges ({\it top}: 1.0$<$$z$$<$1.5; {\it middle}: 1.5$<$$z$$<$2.0; {\it bottom}: 2.0$<$$z$$<$2.5). We also show comparable objects from several recent ALMA studies (\citealt{2015ApJ...806..110D} - black points; \citealt{2016arXiv160706769A} - green points; \citealt{2016arXiv160600227D} - blue points). We overplot the Main Sequence established in each redshift interval by \citet{2014ApJ...795..104W} as solid purple curves, while the dashed curves represent the typical factor of 3 dispersion as suggested in \citet{2015ApJ...806..110D}. 
}
\end{figure}

We also take advantage of the high resolution of \textit{HST} cameras to measure the optical extents of our objects following the method described in \citet{2016ApJ...820...98L}. We initially compare the observed angular sizes measured in the \textit{HST} data with those measured in the ALMA UV-plane 
as these should not rely on assumptions about magnification or redshift. As already discussed, the ALMA fluxes are tracing the cool dust content of the galaxies, which is likely to be more compact than the stellar component traced by {\it HST}. This trend holds true for seven of the 12 ALMA-FFs DSFGs (Fig.~\ref{fig.size}), while the remaining five have comparable sizes. To obtain physical sizes, we fit half light radii in the source plane and calculate errors accounting for both PSF and magnification uncertainties. All of the ALMA-FFs DSFGs appear to be resolved, with a half light radii between $\approx$0.5--2.5\,kpc.

\begin{figure}
\centering
\includegraphics[width=11cm]{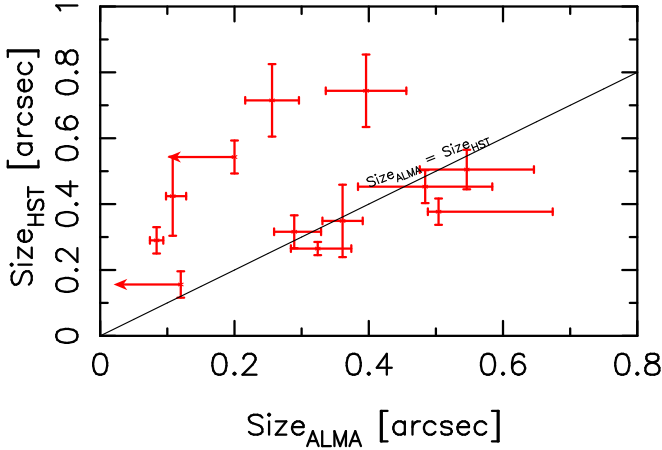}
\caption{\label{fig.size} Comparison of the observed half light radii measured on ALMA maps \citep{Gonzalez2016} and {\it HST} $F140W$/WFC3 images. Assuming that the magnifications are similar at 1.1\,mm and 1.6$\mu$m, more than half of the ALMA-FFs DSFGs show dust emission which is signficantly more compact than the stellar activity probed by {\it HST}, while the rest exhibit comparable extents.}
\end{figure}

All the properties are reported on Table~\ref{tab.properties}.


\begin{table*}
\caption{Physical Properties based on SED analysis}             
\label{tab.properties}      
\centering                          
\begin{tabular}{|l | c | c c c c c | c |c |  }        
\hline\hline                 
ID & $z$ & $\log$[M$_{\star}$] & $\log$[M$_{dust}$]  & $\log$[SFR] & A$_v$ & T$_{dust}$ & Size & $\mu$\\ 
  &     & [M$_{\odot}$] & [M$_{\odot}$] & [M$_{\odot}$/yr] & [mag] & [K] & [kpc]  & \\
\hline
A2744-ID01 & \textit{2.90} &    10.72$^{+ 0.01 }_{-0.10 }$ &    8.34 $^{+0.08}_{-0.04}$ & 1.58$^{+ 0.12}_{-0.05}$ &    3.08 $^{+0.02 }_{- 0.19}$ &    39.87 $^{+ 4.99 }_{-2.02}$ &  1.12 $\pm$ 0.17 & 4.45$^{+0.03}_{-0.05}$ \\
 A2744-ID02 &  \textit{2.48} &    9.98$^{+ 0.10}_{-0.09 }$ &    8.78 $^{+0.01}_{-  0.12}$ &    1.76$^{+ 0.01 }_{-0.08}$ &    2.91$^{+0.04}_{-0.17}$ & 39.55$^{+   2.11 }_{-2.00}$ &  1.66 $\pm$ 0.26 & 2.38 $^{+0.01}_{-0.02}$ \\
 A2744-ID03 &    2.52$^{+  0.23}_{-  0.45}$ &    11.04$^{+  0.13 }_{-  0.09}$ &    8.22$^{+  0.19}_{- 0.07}$ &    1.51$^{+  0.17}_{-  0.30}$ &    2.74$^{+  0.52}_{-  0.28}$ &    40.38$^{+   4.57}_{-   6.33}$ & 2.87$\pm$0.17 & 3.14 $^{+0.09}_{-0.08}$ \\
A2744-ID04 &     1.02$^{+  0.32}_{-  0.09}$ &    10.53     $^{+0.03 }_{-  0.11}$ &    8.40 $^{+  0.19}_{- 0.03 }$ &    1.77$^{+  0.43}_{-  0.19}$  &   3.91$^{+  0.22}_{-  0.73    }$ &    32.02$^{+   5.42 }_{-   1.88 }$ & 2.46$\pm$0.38 & 2.28 $\pm$ 0.06 \\
 A2744-ID05 & 2.01$^{+0.69}_{-0.16}$ &    11.51$^{+   0.04}_{-  0.24}$ & 8.38$^{+  0.13}_{-  0.11}$ &   0.94$^{+  0.28}_{- 0.05}$ &    3.57$^{+  0.48 }_{-  0.66 }$ &    41.60 $^{+   4.85 }_{-   6.65}$ & 4.04 $\pm$ 0.62 & 2.31 $^{+0.23}_{-0.18}$ \\
 A2744-ID06 &    2.08$^{+  0.13}_{-0.08}$ &    11.14 $^{+  0.24}_{- 0.03}$ &    8.38$^{+  0.14}_{-  0.13}$ &    1.29$^{+  0.30}_{- 0.03}$ &    2.97$^{+0.01}_{-  0.41}$ &    44.64$^{+ 1.00 }_{-   10.29}$ & 4.01 $\pm$ 0.62 & 2.62 $^{+0.15}_{-0.13}$ \\
 A2744-ID07 &   1.85 $^{+  0.16}_{-  0.14}$ &    10.86 $^{+  0.07 }_{- 0.08 }$ &    8.23$^{+ 0.09 }_{-  0.14}$ &    1.53 $^{+ 0.03 }_{-  0.38 }$ &    5.17 $^{+ 0.07 }_{-  0.68}$ &    29.45 $^{+   11.30}_{-  3.20}$ &  1.34$\pm$0.24 & 2.44$\pm$0.06 \\
 M0416-ID01 & \textit{2.09} &    10.53 $^{+0.01 }_{- 0.01 }$ &    8.32$^{+ 0.08}_{-  0.16}$ &    2.07$^{+   0.01}_{-0.01}$ &    1.27$^{+ 0.01}_{-0.01}$ &    60.72$^{+   9.63 }_{- 0.01 }$ &   2.00 $\pm$ 0.17 & 1.76$\pm$0.01 \\
 M0416-ID02 &   \textit{1.95}  &    10.74 $^{+0.01}_{- 0.04}$ & 8.14$^{+ 0.08}_{-  0.11}$ &    1.76$^{+0.04 }_{-0.06 }$ &    2.02$^{+0.01}_{-  0.11 }$ &    53.77$^{+ 4.08 }_{-6.12 }$ &  3.31 $\pm$ 0.35 & 2.32 $\pm$ 0.02 \\
 M0416-ID03 &    1.29$^{+  0.11}_{-  0.39}$ &    10.72$^{+ 0.01 }_{- 0.10}$ &    8.34 $^{+   0.08 }_{- 0.04 }$ &   1.58 $^{+  0.12 }_{-  0.05 }$ &   3.08$^{+0.02 }_{-  0.19  }$ &    39.86  $^{+   4.99 }_{-   2.01 }$ & 4.67$\pm$0.37 & 1.55 $\pm$0.02 \\
 M0416-ID04 &    2.27$^{+  0.17}_{-  0.61}$ &    11.15$^{+   0.04}_{-  0.51}$ &    8.13$^{+  0.27 }_{-  0.19}$ &    1.51$^{+  0.49}_{-0.05}$ &    1.75$^{+   1.16}_{- 0.02}$ &    43.91 $^{+   7.14}_{-   8.86}$ &   2.53 $\pm$ 0.29 & 1.93 $^{+0.09}_{-0.12}$ \\
 M1149-ID01 &    \textit{1.46} &    10.91$^{+0.02 }_{-0.18}$ &    8.37$^{+ 0.07}_{-  0.44}$ &    1.49$^{+0.05 }_{-0.03 }$ &    1.24 $^{+0.01 }_{- 0.08 }$ &    37.03$^{+   11.62 }_{-  0.78  }$ &  3.34 $\pm$ 0.42 & 2.51 $^{+0.14}_{-0.10}$ \\

\hline
\end{tabular}
\vspace{0.6cm}
\begin{flushleft}
Columns: (1) ID; (2) Adopted redshifts used with MAGPHYS to estimate physical properties, where italicized values represent spectroscopic redshifts; (3,4,5,6,7) MAGPHYS estimated stellar masses, dust masses, star formation rates, visual extinctions, and dust temperatures, respectively, and their associated 1$\sigma$ errors; (8) size measured on the $F140W$/WFC3 image, with 1$\sigma$ errors including uncertainties from both magnification and PSF; (9) magnification estimates based on the CATS mass models and {\it Lenstool}.\\ 
All properties are corrected for magnification.

\end{flushleft}

\end{table*}



\begin{figure}
\centering
\includegraphics[width=11.cm]{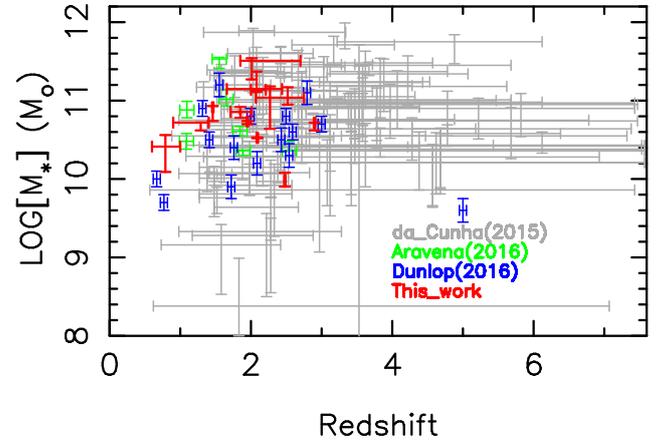}
\caption{\label{fig.M_vs_z} Stellar mass vs. redshift for the ALMA-FFs DSFGs (red), compared to objects from some recent ALMA studies (\citealt{2015ApJ...806..110D} in grey; \citealt{2016arXiv160706769A} in green; and \citealt{2016arXiv160600227D} in blue). The ALMA-FFs survey is probing a relatively wide range of stellar mass at moderate redshift, comparable to other surveys. }
\end{figure}






\subsection{Stacked SED}

Given that the ALMA-FFs DSFGs generally exhibit higher dust attenuation compared to the HUDF and ASPECS samples, a composite spectrum may prove useful for future spectral fitting. To this end, we followed the method described in \citet{2016arXiv160600227D} to compute the stacked SED of all our ALMA 1.1mm sources, taking the mean flux value and the standard deviation as error bar, and assuming that each spectra has the same weight. Thus we de-redshifted all the SEDs to the rest-frame, normalized them to the 1.1mm flux, and then binned the flux per interval of wavelength. The individual and  binned SEDs are shown in Fig.~\ref{fig.stacked.sed}. Interestingly, we do not observe the 8$\mu$m feature in our stacked SED as found in  \citet{2016arXiv160600227D}, although this is may be in part because only half of the sample is observed at 24$\mu$m by {\it Spitzer} and our constraints may be weaker. Importantly, this 8$\mu$m feature looks to be particularly strong only in a few sources within the sample studied by  \citet{2016arXiv160600227D}, and therefore may not be present in all SMGs.

\begin{figure}
\centering
\includegraphics[width=10cm]{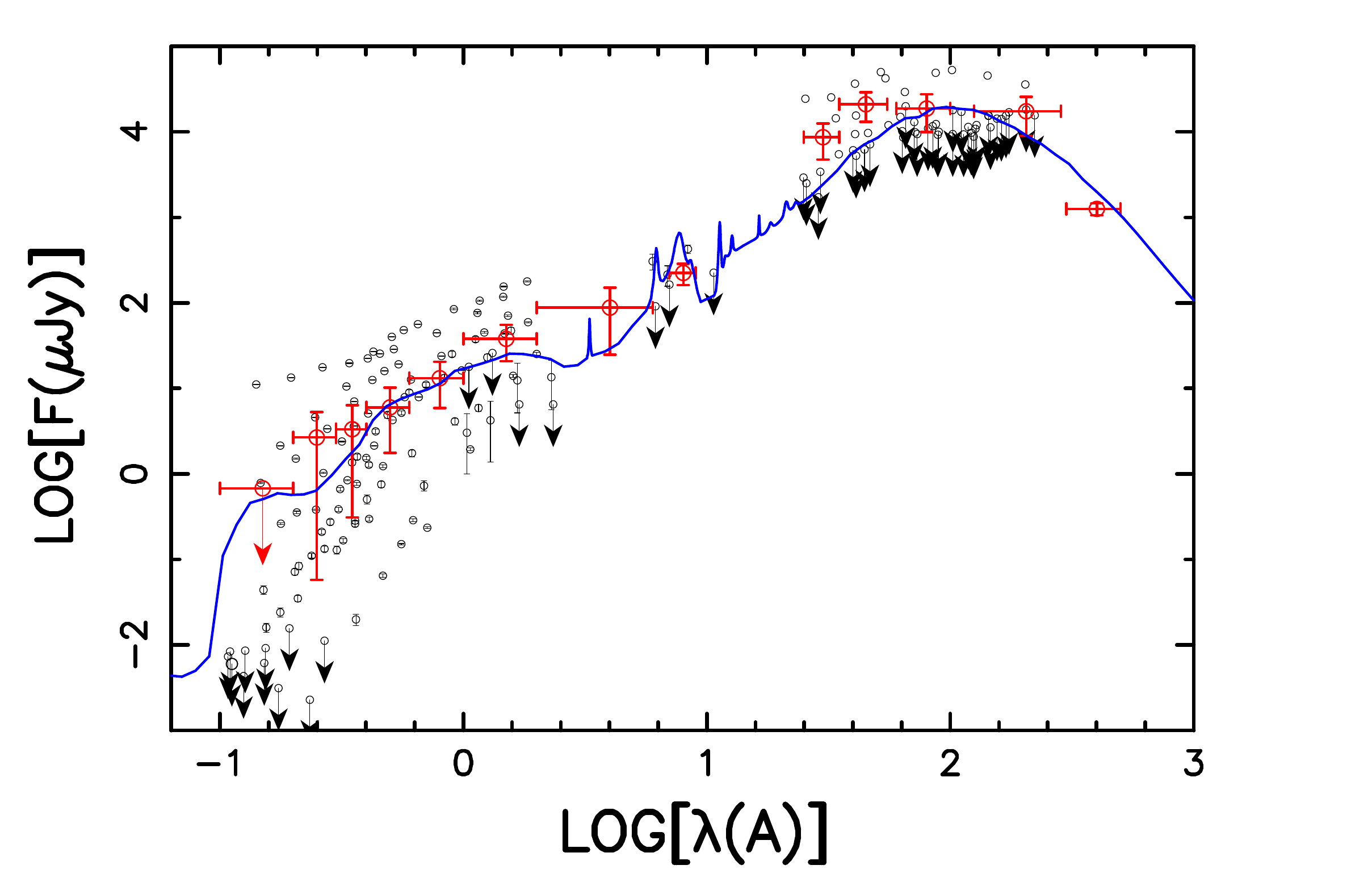}
\caption{\label{fig.stacked.sed} Stacked SED computed from the 12 ALMA 1.1mm sources analyzed in this study. The black dots and limits show the photometry of all the sources after de-redshifting the SEDs and normalizing them by their 1.1\,mm flux density. The red points present the stacked photometry, where error bars on the flux densities are derived from the standard deviation in each bin. The blue curve denotes the SED template from \citet{2015ApJ...814....9K}, combining star formation and AGN contributions.  }
\end{figure}



\section{Summary and Conclusions}

In this paper, we report on the photometric analysis of 12 sources detected at $>$5$\sigma$ on deep ALMA 1.1\,mm maps reaching sensitivities of 55--71\, $\mu$Jy~beam$^{-1}$. We identify optical counterparts for all of our objects thanks to the high-resolution of ALMA and the excellent multi-wavelength coverage available in the FFs. We combine all the data currently available from space (\textit{ HST, Spitzer, Herschel}) and ground-based (\textit{VLT, ALMA}) telescopes in order to constrain their SEDs over a large range of wavelength.  Based on SED-fitting, we estimate several physical properties of our sample, such as redshifts, star formation rates, stellar and dust masses, reddening and dust temperatures and their associated errors. We take advantage of public cluster mass models released in the framework of the FFs survey to estimate the light amplification for our objects, and therefore correct all physical parameters for lensing effects. 

As already demonstrated by several previous studies, photometric redshift estimation for sub-mm galaxies can benefit from using a wide array of SED templates covering the full UV to radio wavelength range (of which there are not so many). In order to get reliable photometric redshifts, we estimate the photometric redshifts using only optical/NIR SEDs, only FIR SEDs, and the combined optical/NIR/FIR SEDs. We used a large range of templates covering optical to sub-mm wavelengths, for several different types of objects (AGN, ULIRGS, etc...). For several sources, the photometric redshift estimates from the NIR-SED and the FIR-SED were not consistent. In general, we adopte the redshift estimate for which the reduced $\chi^2_{\nu}$ was minimized. However, the number of photometric constraints in the NIR for two objects was not sufficient to draw any firm conclusions and NIR+FIR or FIR SED estimates were adopted. We also take advantage of the published GLASS spectroscopic survey to obtain spectroscopic redshifts for $\approx$40\% of our sample. Reassuringly, our photometric redshift estimates 
were in line with these spectroscopic redshifts for all objects, confirming the method we used to estimate the photo-$z$. The redshift range for the ALMA-FFs DSFGs was $z$$=$1.0--2.9, with a mean of 1.99$\pm$0.27, that is in perfect agreement with what has been predicted using phenomenological models. A better estimation of the FIR photometry, especially in terms of deblending, would strongly improve the quality of the photometric redshift estimates, and therefore could be used to build new templates combining optical, NIR and FIR photometry. 

We use MAGPHYS, a public SED analysis tool, to estimate the physical properties of our sample. The best-fitted SEDs imply that the ALMA-FFs DSFGs trace relatively massive star-forming galaxies with 10.0$<\log[M_{\star}/M^{\odot}]<$11.5 and 0.9$<\log[SFR/(M^{\odot} yr^{-1})]<$2.0. These values are in good agreement with previous ALMA studies, for example the 99 SMGs from the ALESS survey have a mean stellar mass of 10.95$^{+0.6}_{-0.8}$, where we found <$\log$M$_{\star}$>=10.82$\pm$0.40. The sub-mJy ALMA DSFGs still appear to  probe relatively massive galaxies ($\log[M_{\star}/M_{\odot}]\gtrsim10$), albeit with lower SFRs. We demonstrate that ALMA observations are probing a more compact region that the stellar activity explored by \textit{HST}. Their morphologies and colors indicate that these galaxies remain distinct from the Ly-$\alpha$ emitter and Lyman Break Galaxy populations at similar redshifts, and demonstrate that the strong division between the UV and submm/mm selected objects remains even at these faint fluxes and more modest SFRs where the two populations share considerable overlap. 

Robust spectroscopic confirmation of the ALMA-FFs redshifts, either in the NIR or at mm wavelengths, will be crucial to pin down their nature and properties better, as well as trace molecular lines that can be used for a variety of star formation and kinematic diagnostics. The ALMA-FFs sources are factors of a few brighter than those observed in the HUDFs, and thus should be easier to follow up. 


\begin{acknowledgements}
We would like to thank the anonymous referee for the constructive comments we received to improve our paper.
We acknowledge support from:
European Research Council Advanced Grant FP7/669253 (NL, RSE); CONICYT-Chile grants Basal-CATA PFB-06/2007 (NL, FEB, PT-I, JG-L, SK, RC, RD, LI, NN, NP, CR-C, ET), CONICYT-Gemini 32120003 (NL, JG-L), FONDECYT Regular 1141218 (NL, FEB, JG-L) and 1130528 (ET), FONDECYT Postdoctorado 3140542 (PT-I), 3150238 (CR-C) and "EMBIGGEN Anillo ACT1101 (FEB, NN, ET); the Ministry of Economy, Development, and Tourism's Millennium Science Initiative through grant IC120009, awarded to The Millennium Institute of Astrophysics, MAS (FEB, TA, CR-C). M.A. acknowledges partial support from FONDECYT through grant 1140099. RP thanks the support of the OCEVU Labex (ANR-11-LABX-0060) and the A*MIDEX project (ANR-11-IDEX-0001-02) funded by the "Investissements d'Avenir" French government program managed by the ANR.
This paper makes use of the following ALMA data: ADS/JAO.ALMA\#2013.1.00999.S. ALMA is a partnership of ESO (representing its member states), NSF (USA) and NINS (Japan), together with NRC (Canada) and NSC and ASIAA (Taiwan), in cooperation with the Republic of Chile. The Joint ALMA Observatory is operated by ESO, AUI/NRAO and NAOJ. This work is based on observations made with the NASA/ESA Hubble Space Telescope, obtained at the Space Telescope Science Institute (STScI), which is operated by the Association of Universities for Research in Astronomy, Inc., under NASA contract NAS 5-26555. The HST image mosaics were produced by the Frontier Fields Science Data Products Team at STScI. This work is based in part on observations made with the \textit{Spitzer} Space Telescope, which is operated by the Jet Propulsion Laboratory, California Institute of Technology under a contract with NASA.
This work utilizes gravitational lensing models produced by PIs Bradac, Ebeling, Merten \& Zitrin, Sharon, and Williams funded as part of the HST Frontier Fields program conducted by STScI. STScI is operated by the Association of Universities for Research in Astronomy, Inc. under NASA contract NAS 5-26555. The lens models were obtained from the Mikulski Archive for Space Telescopes (MAST).
\end{acknowledgements}
\bibliographystyle{aa}  
\bibliography{ALMA2.bib} 

\clearpage

\begin{landscape}
\begin{table}
\scriptsize
\caption{Photometry of the 5$\sigma$ ALMA detected sources combining \textit{HST}, \textit{VLT}, \textit{Spitzer}  and \textit{Herschel} data.}             
\label{tab.photometry}      
\begin{tabular}{l|ccccccc|c|ccccc|ccccc|}       
\hline\hline                 

ID   & F435W  & F606W & F814W & F105W & F125W & F140W & F160W & K$_s$ & 3.6$\mu$m & 4.5$\mu$m & 5.8$\mu$m & 8.0$\mu$m & 24$\mu$m & 100$\mu$m & 160$\mu$m & 250$\mu$m & 350$\mu$m & 500$\mu$m  \\
      &         &   &   &   &   &   &   &    &   &   &   &   &   &   &      &   &  &    \\ 
 A2744-ID01  &  $>$29.04 &  28.24  &  27.52  &  26.41  &  25.63  &  25.01  &  24.44  & 23.07 &   22.22  &   21.83   &  \textit{blended}   &  \textit{blended}  &   $>$18.7  & $>$15.4  &   $>$14.5  &   $>$13.8  &   $>$13.4   &  13.39   \\
                        &     -            &  $\pm$0.54 &  $\pm$0.14  &  $\pm$0.13  &  $\pm$0.12  &  $\pm$0.14  &  $\pm$0.12  & $\pm$0.11  &   $\pm$0.05  &   $\pm$0.05   &  -   &  -  &   -   &   -  &   -  &   -  &   -   &  $\pm$0.26  \\
 A2744-ID02  &  $>$28.49 &  $>$29.75  &  $>$ 30.09  &  26.09  &  25.32  &  24.61  &  24.17  & 24.21  &   22.66  &   22.30   &  21.13   &  21.04  &   $>$18.3  &   $>$15.4  &   $>$14.5  &   $>$13.7  &   $>$13.3   &  $>$13.6   \\
                        &     -            &  -                &  -                 &  $\pm$0.13  &  $\pm$0.12  &  $\pm$0.14  &  $\pm$0.12  & $\pm$0.15  &   $\pm$0.50 &   $\pm$0.50   &  $\pm$0.50   &  $\pm$0.50  &   -   &  - &  - &  -  & - & -    \\
 A2744-ID03  &  $>$28.67 &  27.34  &  27.38  &  25.80  &  24.73  &  24.17  &  23.57  & 22.46  &   $>$ 21.1  &   $>$20.7   &  $>$22.2   &  $>$22.2  &  $>$18.3   &   $>$15.4  &   $>$14.5  &   $>$13.6  &   $>$13.2   &  $>$13.4   \\
                        &     -            &  $\pm$0.22  &  $\pm$0.14  &  $\pm$0.13  &  $\pm$0.13  &  $\pm$0.12  &  $\pm$0.14  & $\pm$0.06  &   -  &   -   &  -   &  -  &   -   &   -  &   -  &   -  &   -   &  -    \\
 A2744-ID04  & $>$28.51  &  $>$27.8   &  23.99 &  22.93  &  22.49  &  22.24  &  21.97  & 21.28  &   20.63  &   20.25   &  \textit{blended}   &  \textit{blended}  &   18.26   &   13.38  &   12.36  &   12.76  &   12.53   &  $>$12.92   \\
                        &     -            &  -               &  $\pm$0.13  &  $\pm$0.13  &  $\pm$0.12  &  $\pm$0.14  &  $\pm$0.12  & $\pm$0.09  &   $\pm$0.05  &   $\pm$0.05   &  -   &  -  &  $\pm$0.36  & $\pm$0.06	 &   $\pm$0.05  &   $\pm$0.21 &   $\pm$0.30    &  -  \\
 A2744-ID05  &  $>$28.46 &  $>$28.43  &  $>$28.76  &  26.65  &  24.84  &  24.06  &  23.57  & 22.38  &   21.10  &   20.67   &  \textit{blended}   &  \textit{blended} &   $>$18.26   &   $>$15.4  &   $>$14.5  &   $>$13.8  &   $>$13.4   &  $>$13.7   \\
                        &     -            &  -  &  -  &  $\pm$0.13  &  $\pm$0.12 &  $\pm$0.14  &  $\pm$0.12  & $\pm$0.08  &   $\pm$0.05  &   $\pm$0.05   &  -  &  -  &   -   &   -  &   -  &   -  &   -   &  -   \\
 A2744-ID06  &  $>$29.81 &  26.75  &  25.77  &  23.57  &  22.81  &  22.39  &  22.08  & 21.06  &   20.65  &   19.87   &  \textit{blended}   &  \textit{blended}  &  17.54   &   14.74  &   14.24  &   12.32  &   $>$12.61   &  $>$12.95   \\
                        &     -            &  $\pm$0.14  &  $\pm$0.22  &  $\pm$0.30  &  $\pm$0.12  &  $\pm$0.14  &  $\pm$0.12  & $\pm$0.08  &   $\pm$0.05  &   $\pm$0.05   & -   &  -  &   $\pm$0.19   &   $\pm$0.20  &    $\pm$0.28 & $\pm$0.14 &   -   &  -  \\
 A2744-ID07  &  $>$29.81 &  $>$29.76  &  $>$29.59  &  27.54  &  26.45  &  25.89  &  25.46  & 24.03  &   22.13  &   21.49   &  \textit{blended}   &  \textit{blended}  &   18.70   &   $>$15.4  &   $>$14.5  &   $>$13.5  &   $>$13.3   &  $>$13.9   \\
                        &     -            &  -                &  -  &  $\pm$0.15  &  $\pm$0.12  &  $\pm$0.14  &  $\pm$0.12 & $\pm$0.20 &   $\pm$0.05  &   $\pm$0.05   &  -   &  -  &   $\pm$0.36   &   -  &   -  &   -  &   -   &  -   \\
\hline
M0416-ID01   &   22.72   &   22.53   &   22.22   &   22.09   &   21.86   &   21.65   &   21.46   &   20.82   &   20.06   &   19.95   &   -   &  -  &   -   &   14.12  &   13.41  &   $>$13.6  &   $>$13.4   &  $>$13.4    \\
              & $\pm$0.12   & $\pm$0.12   & $\pm$0.12   & $\pm$0.12   & $\pm$0.12  & $\pm$0.12  & $\pm$0.12  &$\pm$0.05   &$\pm$0.05   &$\pm$0.05 &   -   &  -  &   -   &   $\pm$0.20  &   $\pm$0.23  &   -  &   -   &  -   \\
M0416-ID02   &   26.15   &   24.96   &   24.20   &   23.17   &   22.52   &   22.23   &   22.01   &   21.12   &   21.04   &   20.55   &   -   &  - &   -  &   14.78  &   13.67  &   $>$13.6  &   $>$13.4   &  $>$13.5   \\
              & $\pm$0.12   & $\pm$0.12   & $\pm$0.12   & $\pm$0.12  & $\pm$0.12   &  $\pm$0.12 & $\pm$0.12   &$\pm$0.05   &$\pm$0.05   &$\pm$0.05 &   -   &  -  &   -   &   $\pm$0.36  &   $\pm$0.29  &   -  &   -   &  -   \\
M0416-ID03   &   $>$29.30   &   $>$29.20   &   27.02   &   24.25   &   23.67   &   23.35  &   23.22   &   21.89   &   20.52   &   20.63   &   -   &  -  &   -  &   14.39  &   $>$13.9  &   $>$13.5  &   $>$13.4   &  $>$13.6  \\
              &    -   &   -   & $\pm$0.12  & $\pm$0.12  & $\pm$0.12   & $\pm$0.12   & $\pm$0.12   &$\pm$0.05   &$\pm$0.05   &$\pm$0.05 &   -   &  -  &  -   &   $\pm$0.25  &   - &   -  &   -   &  -    \\
M0416-ID04   &   26.69   &   26.15  &   25.37   &   24.49   &   23.69   &   23.32   &   23.01   &   21.86   &   21.19   &   20.86   &   -   &  -  &   -   &   $>$14.8  &   $>$13.9  &   $>$13.5  &   $>$13.4   &  $>$13.6   \\
              & $\pm$0.12   & $\pm$0.12   & $\pm$0.12  & $\pm$0.12   & $\pm$0.12  & $\pm$0.12  & $\pm$0.12  &$\pm$0.05   &$\pm$0.05   &$\pm$0.05 &   -   & -  &  -   &   -  &   -  &   -  &   -   &  -   \\
\hline
.M1149-ID01   &   24.21   &   23.38   &   22.48   &   21.46   &   21.03   &   20.83   &   20.66 & -      &   19.56   &   19.41   &  -   &  -  &  -  &   15.21  &   14.13  &   $>$13.3  &   $>$13.4   &  $>$13.7   \\
              &  $\pm$0.12  &  $\pm$0.12   &  $\pm$0.12  &  $\pm$0.12   &  $\pm$0.12   &  $\pm$0.12 &   $\pm$0.12  & - & $\pm$ 0.05   & $\pm$ 0.05 &  -   &  -  &   -   &   $\pm$0.30  & $\pm$0.26 & - & -  & -   \\

\end{tabular}\\
\vspace{0.3cm}
AB magnitude photometry for the ALMA-FF sources. All the magnitudes have been aperture-corrected, are the observed magnitudes and represent the total flux associated with the source. \\
Columns: (1) ID; \\
(2,3,4,5,6,7,8) \textit{HST} magnitudes measured on \textit{Frontier Fields} images, with non-detections listed as 1$\sigma$  upper limits; \\
(9) HAWK-I magnitudes; \\
 (10, 11, 12, 13, 14) \textit{IRAC} and {MIPS} magnitudes, with non-detections listed as 1$\sigma$ upper limits --- if deblending was not possible, we adopt as an upper limit the flux measured at the position of the object; \\
 (15, 16, 17, 18, 19) \textit{Herschel} magnitudes, with non-detections listed as 3$\sigma$ upper limits, estimated by measuring the flux at the position of our objects. \\
\end{table}

\end{landscape}

\clearpage
\begin{figure*}
\includegraphics[width=9.1cm]{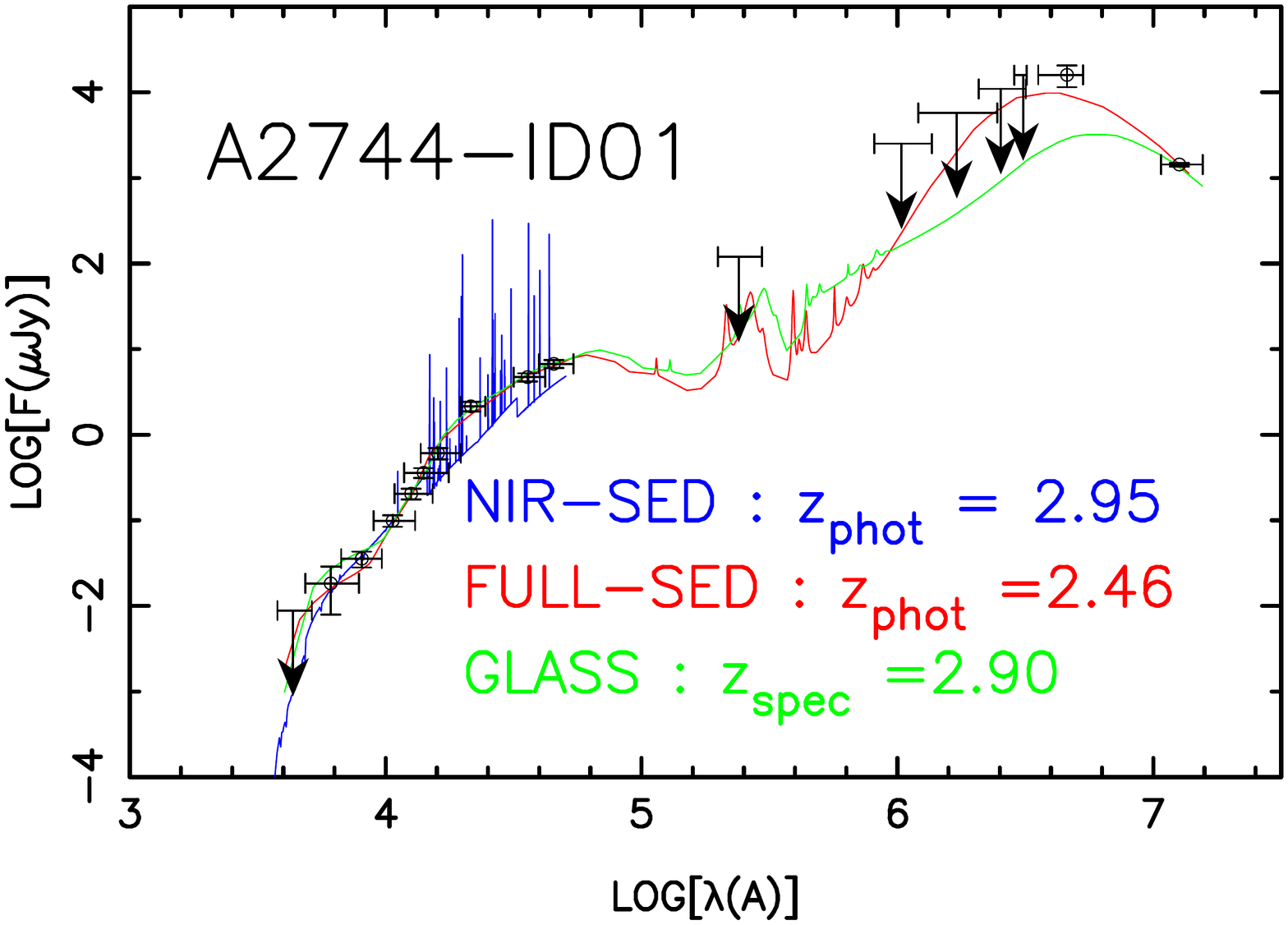}
\includegraphics[width=9.1cm]{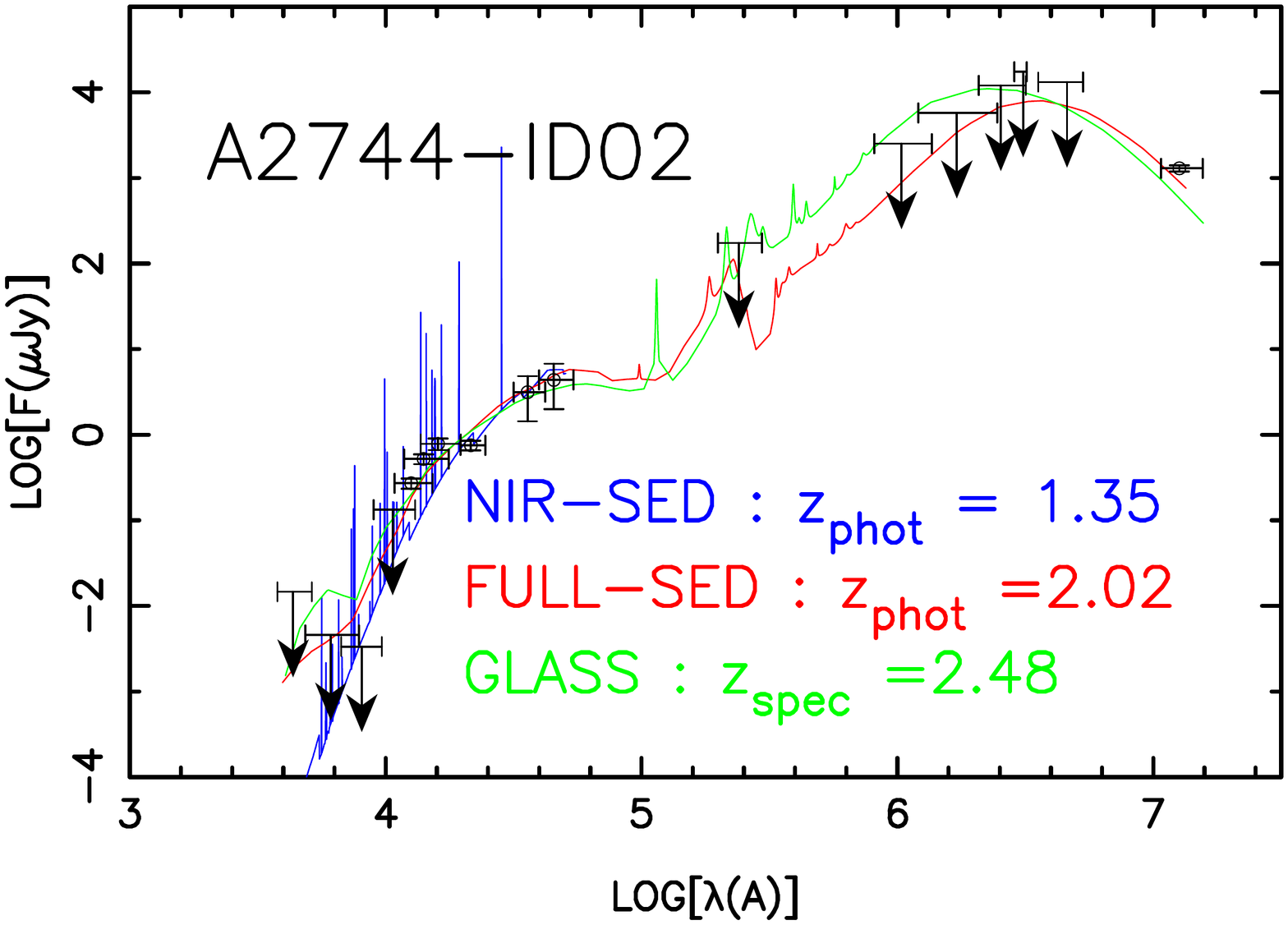} \\
\includegraphics[width=9.1cm]{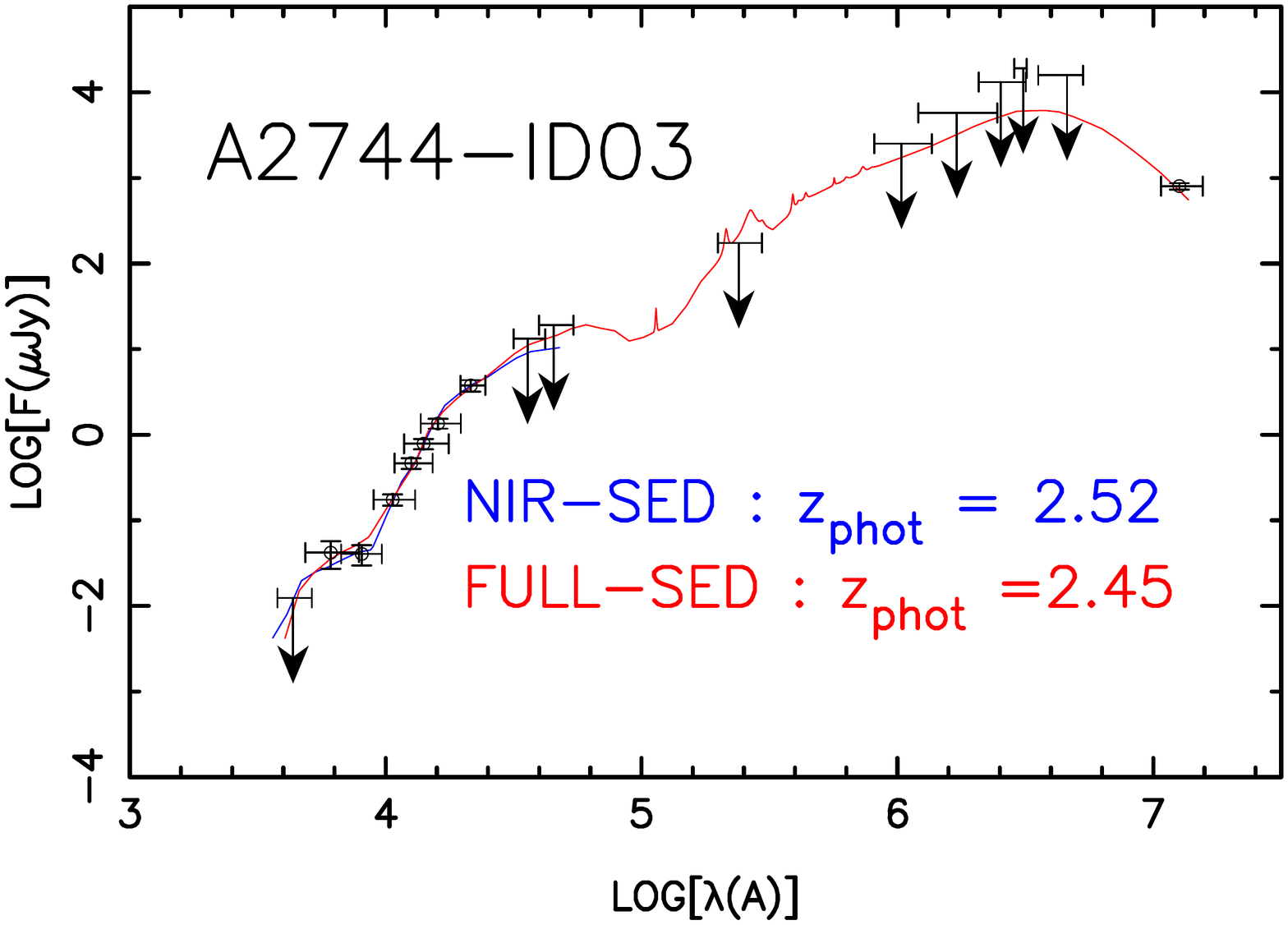}
\includegraphics[width=9.1cm]{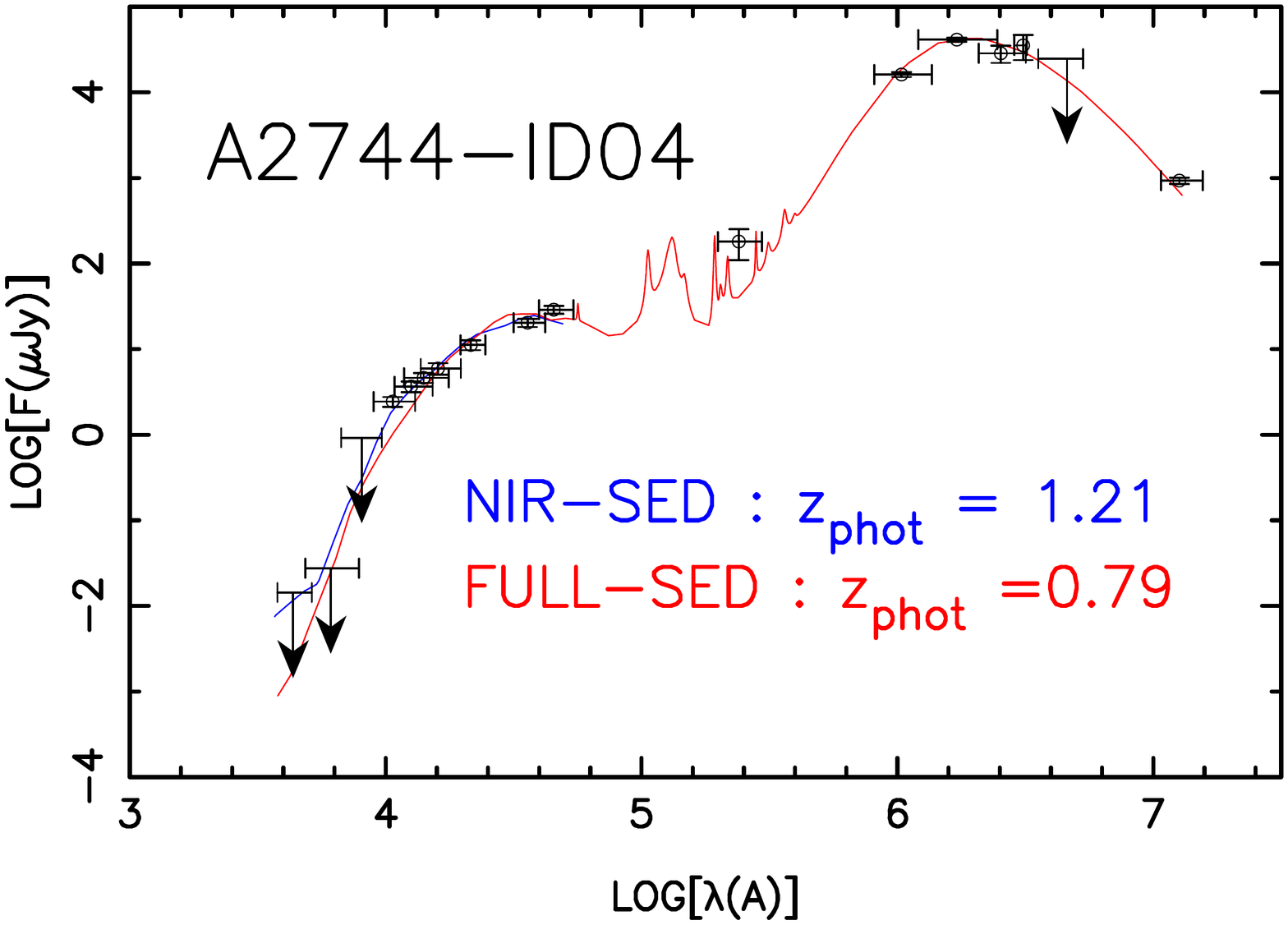} \\
\includegraphics[width=9.1cm]{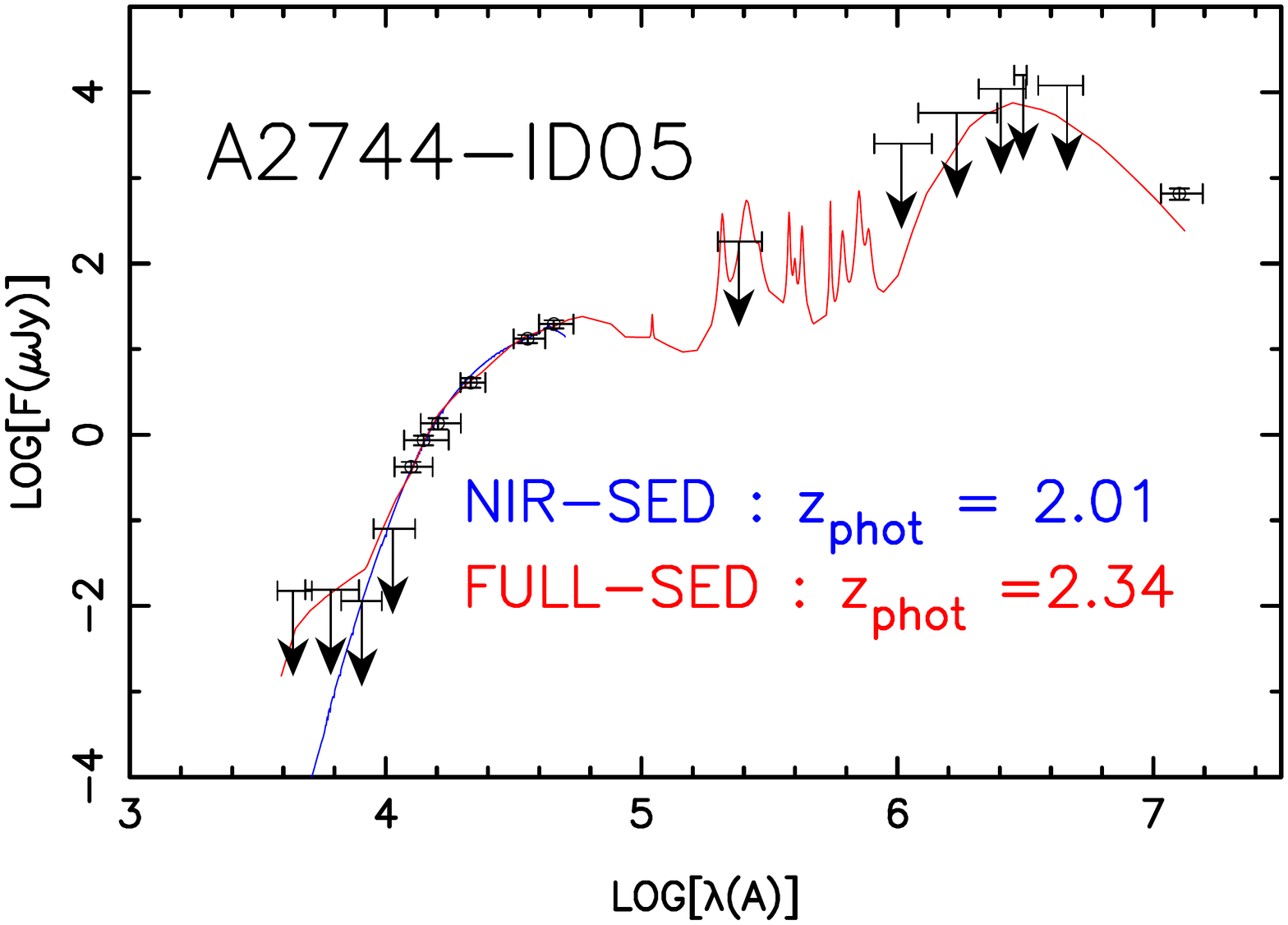}
\includegraphics[width=9.1cm]{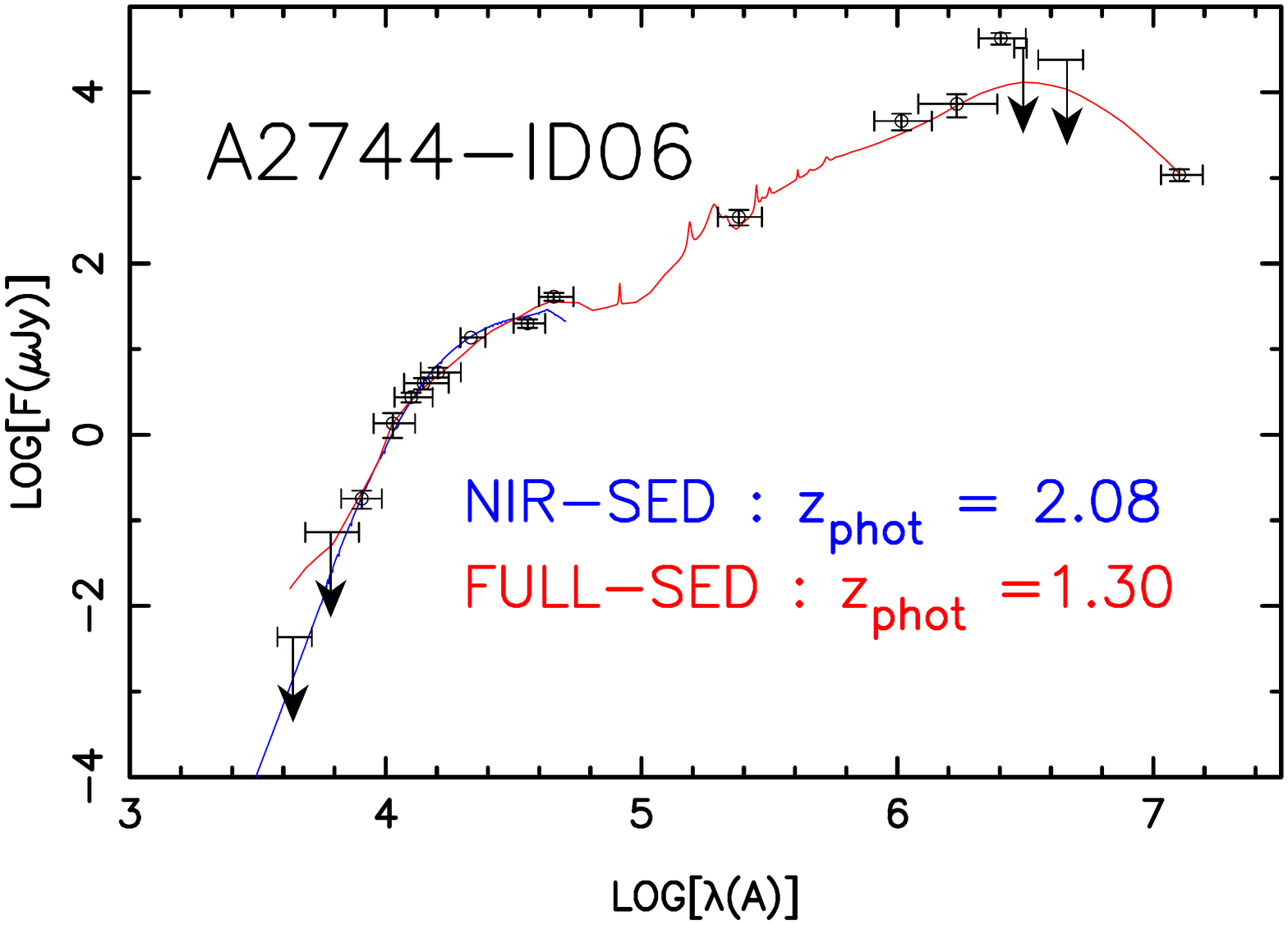} \\
\caption{\label{fig.SED}SED-fitting of ALMA 1.1mm sources optical counterpart having a robust measurement of their redshift. For each source, we plot the best fit of the NIR-SED (blue), the full SED (using upper limits in the FIR, red) and the best SED-fit when the redshift is fixed to the spectroscopic redshift when available (green). 
  }
\end{figure*}

\begin{figure*}
\ContinuedFloat
\includegraphics[width=9.1cm]{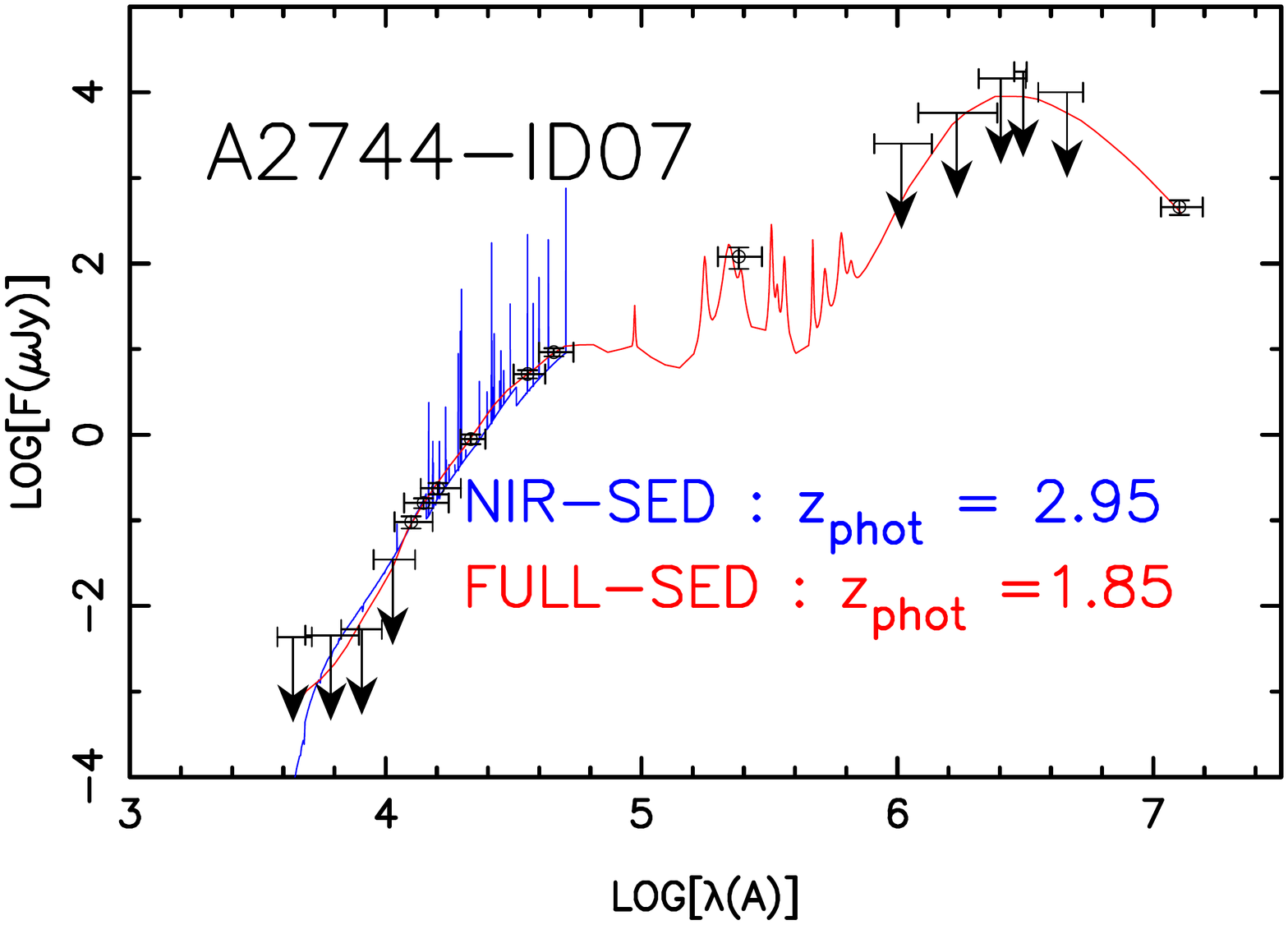}
\includegraphics[width=9.1cm]{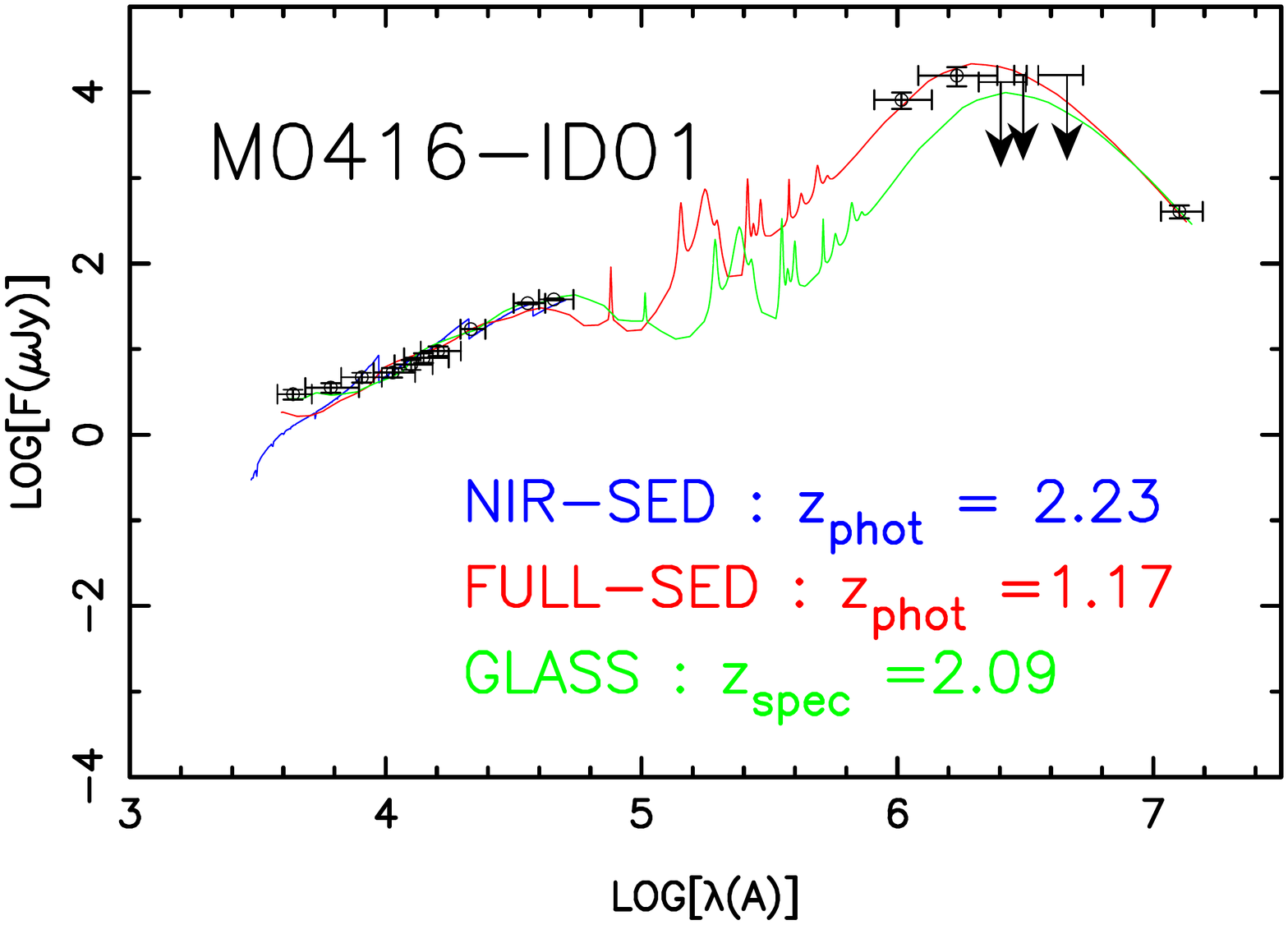}\\
\includegraphics[width=9.1cm]{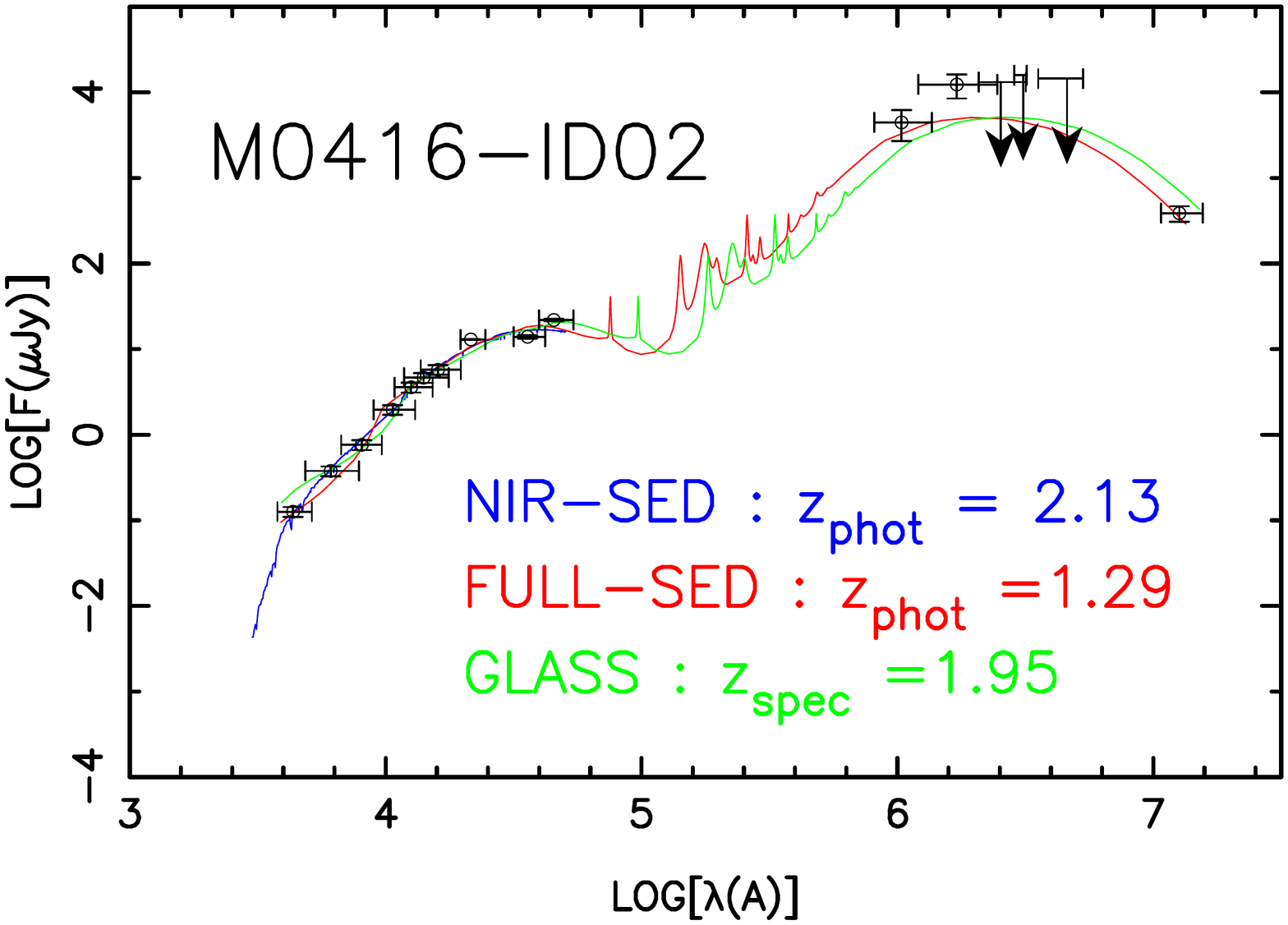} 
\includegraphics[width=9.1cm]{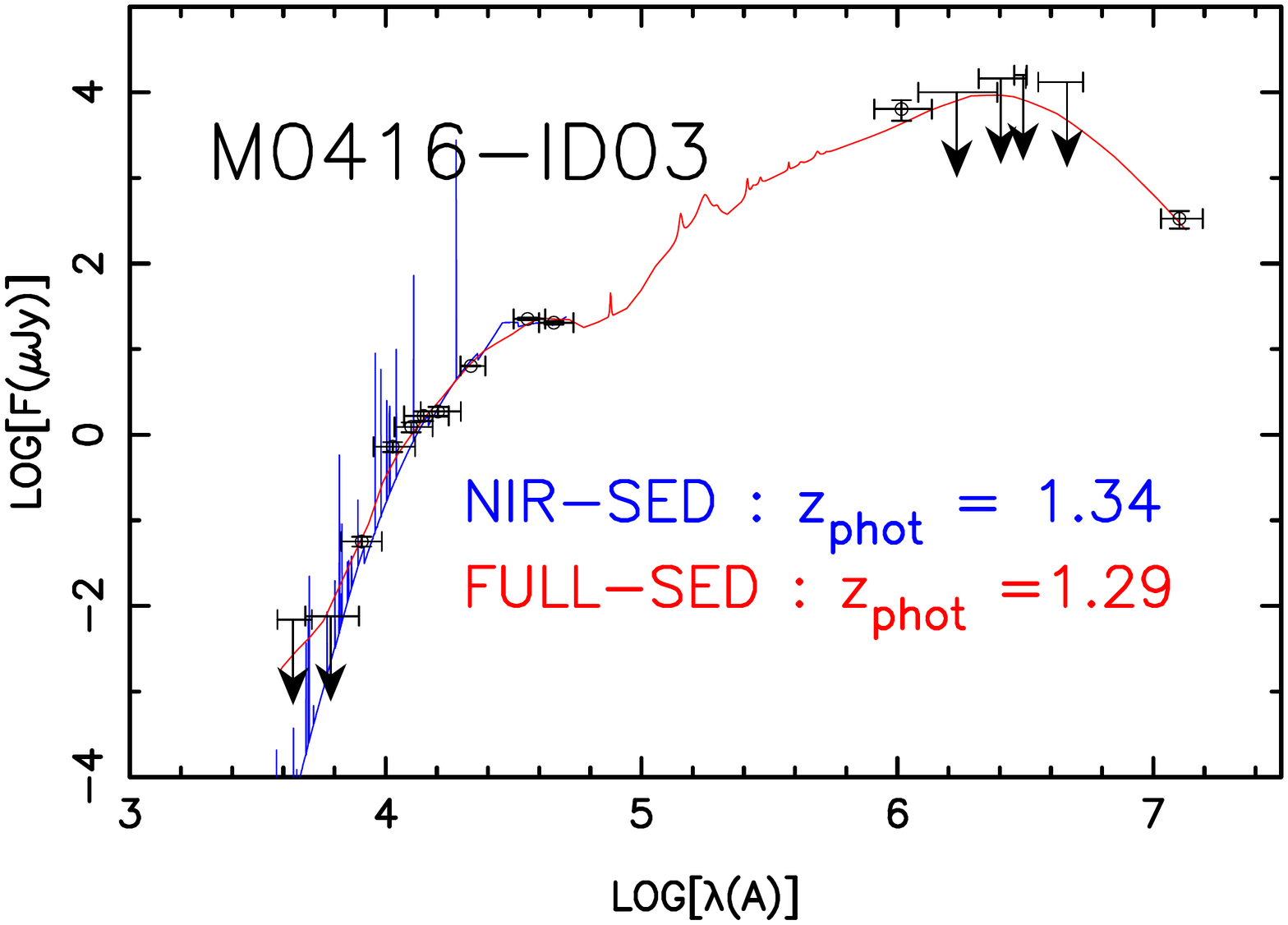}\\
\includegraphics[width=9.1cm]{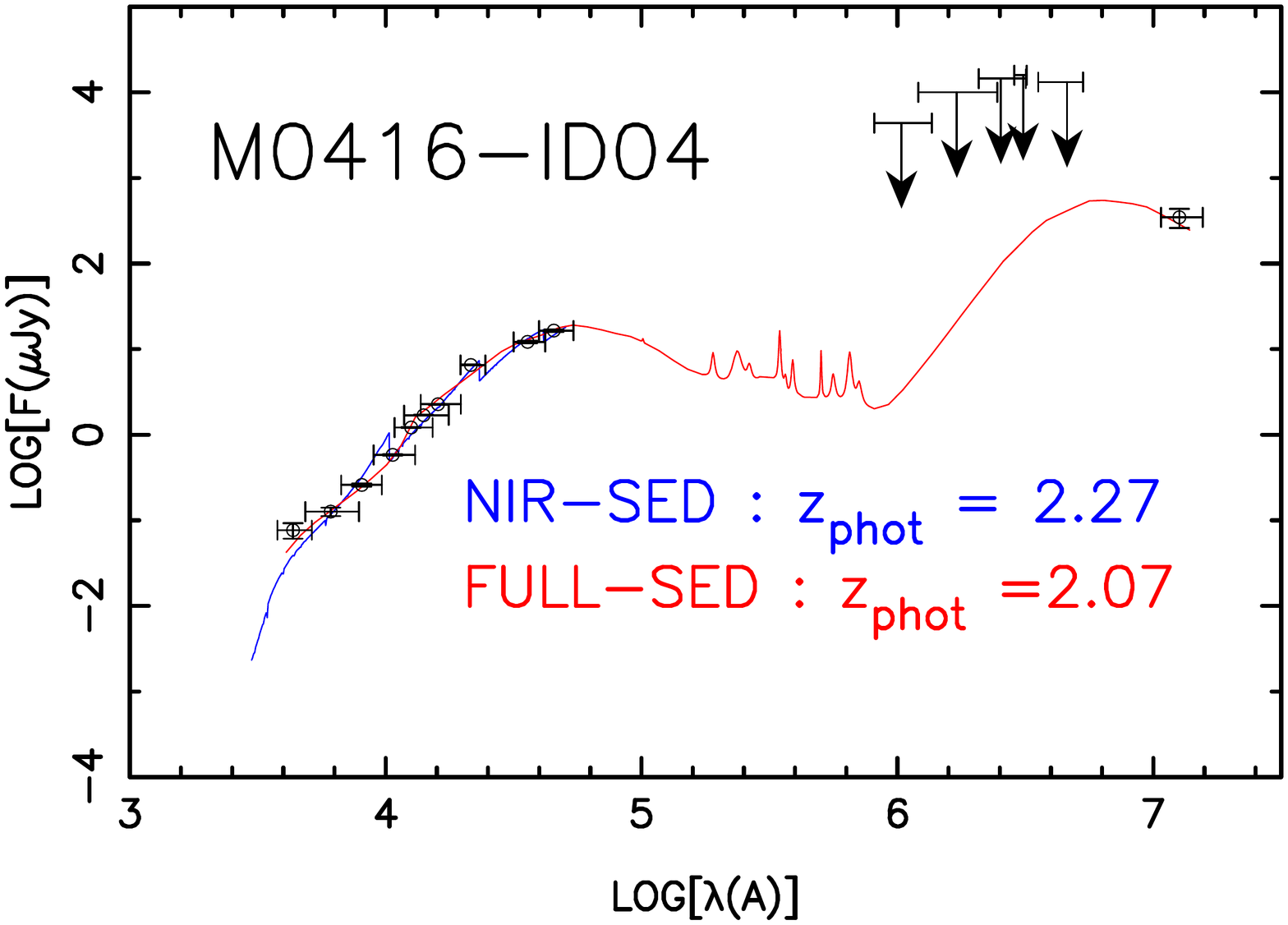} 
\includegraphics[width=9.1cm]{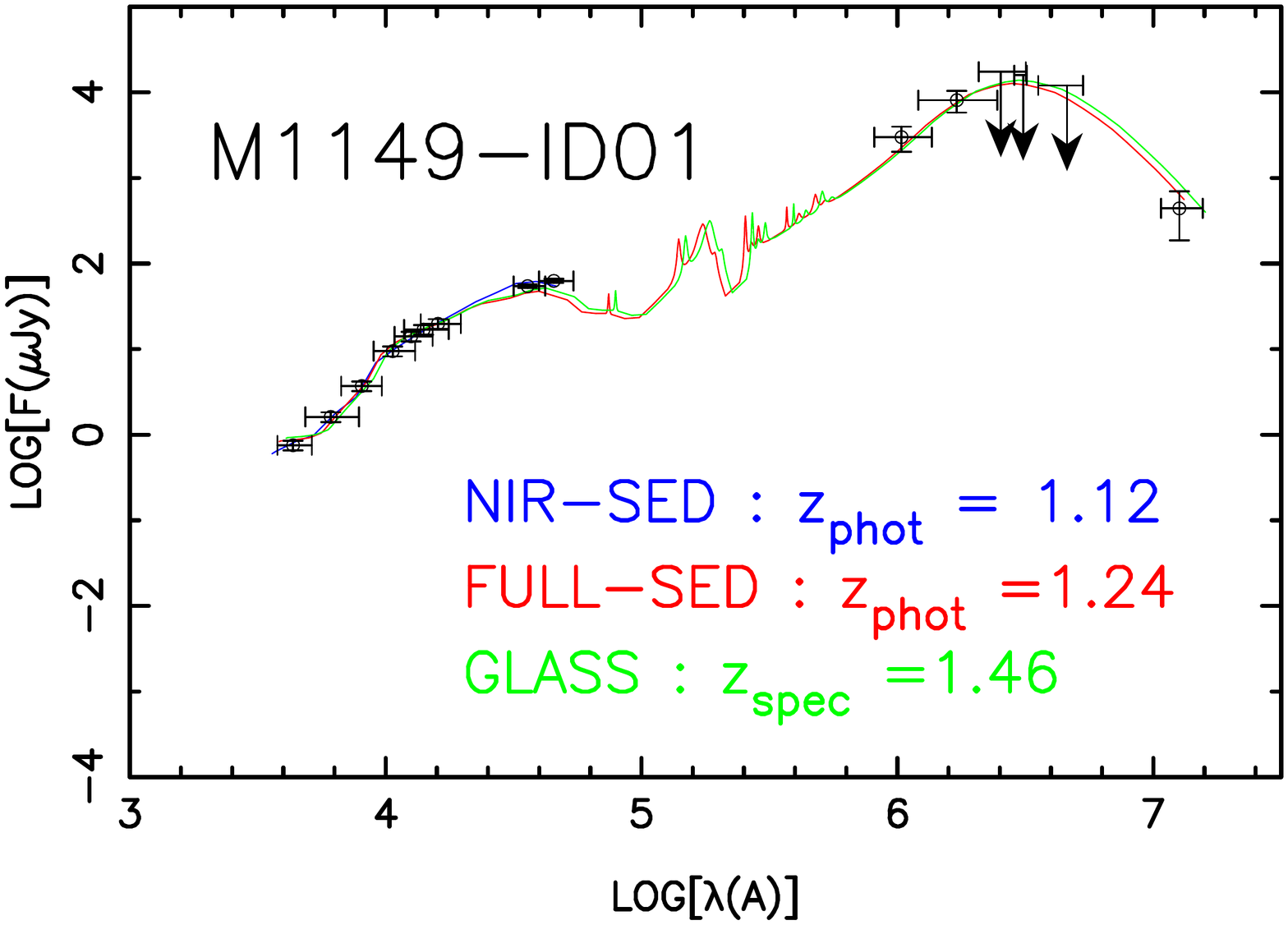} 
\caption{same as previous page}
\end{figure*}




\end{document}